\documentclass[aps,pra,twocolumn,superscriptaddress]{revtex4-1} 

\usepackage{amsmath,amssymb,amsfonts,bm,latexsym}

\usepackage[pdftex]{graphicx,color}
\usepackage{hyperref}
\hypersetup{
   colorlinks=true,        
   linkcolor=blue,          
   citecolor=blue,        
  filecolor=magenta,      
   urlcolor=blue           
}

\newcommand{\bra}[1]{\langle #1|}
\newcommand{\ket}[1]{|#1 \rangle}
\newcommand{\bracket}[2]{\langle #1|#2 \rangle}

\newcommand{\rv}{{\mathbf{r}}}

\newcommand{\kv}{{\mathbf{k}}}

\newcommand{\vv}{{\mathbf{v}}}

\newcommand{\zerov}{{\mathbf{0}}}

\newcommand{\Av}{{\mathbf{A}}}

\newcommand{\Omegat}{\tilde{\Omega}}

\newcommand{\Lcal}{\mathcal{L}}

\newcommand{\ua}{\uparrow}
\newcommand{\da}{\downarrow}

\newcommand{\gt}{\tilde{g}}

\newcommand{\erm}{\mathrm{e}}
\newcommand{\Tr}{\mathrm{Tr}}
\newcommand{\osc}{\mathrm{osc}}
\newcommand{\zero}{\mathrm{zero}}

\newcommand{\drm}{\mathrm{d}}
\newcommand{\irm}{\mathrm{i}}
\newcommand{\srm}{\mathrm{s}}
\newcommand{\Brm}{\mathrm{B}}

\begin{document}
\title{
Intercomponent entanglement entropy and spectrum \\
in binary Bose-Einstein condensates
}
\author{Takumi Yoshino}
\email{yoshino@cat.phys.s.u-tokyo.ac.jp}
\affiliation{Department of Physics, University of Tokyo, 7-3-1 Hongo, Bunkyo-ku, Tokyo 113-0033, Japan}
\author{Shunsuke Furukawa}
\email{furukawa@rk.phys.keio.ac.jp}
\affiliation{Department of Physics, 
Keio University, 3-14-1 Hiyoshi, Kohoku-ku, Yokohama 223-8522, Japan}
\author{Masahito Ueda}
\affiliation{Department of Physics, University of Tokyo, 7-3-1 Hongo, Bunkyo-ku, Tokyo 113-0033, Japan}
\affiliation{RIKEN Center for Emergent Matter Science (CEMS), Wako, Saitama 351-0198, Japan}
\affiliation{Institute for Physics of Intelligence, University of Tokyo, 7-3-1 Hongo, Bunkyo-ku, Tokyo 113-0033, Japan}
\date{\today}
\begin{abstract}
We study the entanglement entropy and spectrum between components in binary Bose-Einstein condensates in $d$ spatial dimensions. 
We employ effective field theory to show that the entanglement spectrum exhibits an anomalous square-root dispersion relation 
in the presence of an intercomponent tunneling (a Rabi coupling) and a gapped dispersion relation in its absence. 
These spectral features are associated with the emergence of long-range interactions in terms of the superfluid velocity and the particle density in the entanglement Hamiltonian. 
Our results demonstrate that unusual long-range interactions can be emulated in a subsystem of multicomponent BECs that have only short-range interactions. 
We also find that for a finite Rabi coupling the entanglement entropy exhibits a volume-law scaling with subleading logarithmic corrections 
originating from the Nambu-Goldstone mode and the symmetry restoration for a finite volume. 
\end{abstract}
\maketitle


\section{Introduction}

Over the last two decades, the concept of quantum entanglement has been extensively applied to quantum many-body problems \cite{Laflorencie16,Calabrese09_ex,Amico08}. 
A useful measure of entanglement for a many-body state $|\Psi\rangle$ is the entanglement entropy (EE). 
By partitioning a many-body system into a subregion $A$ and its complement $\bar{A}$, 
the EE is defined as the von Neumann entropy $S_A=-\Tr~ \rho_A \ln \rho_A$ of the reduced density matrix $\rho_A = \Tr_{\bar{A}} |\Psi\rangle \langle\Psi |$. 
The dependence of $S_A$ on the size of $A$ can exhibit a universal scaling that reflects the long-distance properties of the system. 
When the ground state $|\Psi\rangle$ contains only short-range correlations, the EE scales with the size of the boundary of $A$ (boundary law) \cite{Srednicki93,Eisert10}. 
The deviation from a boundary law signals the presence of certain nontrivial correlations. 
In one-dimensional (1D) quantum critical systems, for example, the EE exhibits a universal logarithmic scaling 
with a coefficient determined by the central charge of the underlying conformal field theory \cite{Holzhey94, Vidal03, Calabrese04,Calabrese09}. 
In 2D topologically ordered systems, 
the EE obeys a boundary law, but there appears a universal subleading constant that reflects the underlying topological order \cite{Kitaev06, Levin06, Hamma05, Hamma05PhysLettA}. 

More detailed information about bipartite entanglement can be investigated by using the entanglement spectrum (ES) \cite{Li08}. 
By rewriting the reduced density matrix in the form of $\rho_A=\erm^{-H_\erm}$, where $H_\erm$ is referred to as the entanglement Hamiltonian, 
the ES is defined as the full eigenvalue spectrum of $H_\erm$. 
The EE then corresponds to the thermal entropy in the canonical ensemble given by $H_\erm$ at the fictitious temperature $T=1$. 
The ES finds particularly useful applications in gapped topological phases---the ES has been found 
to exhibit the same low-energy features as the edge-mode spectrum in such diverse examples as
quantum Hall states \cite{Li08, Thomale10FQH, Sterdyniak12, Dubail12FQH, Rodriguez12, Rodriguez13, Yan19}, 
fractional Chern insulators \cite{Regnault11}, topological insulators \cite{Turner10, Fidkowski10, Alexandradinata11,Fang13}, 
and symmetry protected topological phases \cite{Pollmann10, Turner11, Fidkowski11, Cho17}. 
Several physical arguments have been put forth in support of this remarkable edge-entanglement correspondence 
\cite{Qi12,Chandran11,Dubail12proof,SwingleSenthil12}. 
Among them, the argument by Qi, Katsura, and Ludwig \cite{Qi12} is based on the following illuminating idea, which is referred to as the ``cut and glue'' picture (see also Refs.~\cite{Lundgren13,Cano15,Das15,WenX16,Sohal20}). 
Suppose we cut a 2D topological state into two pieces $A$ and $\bar{A}$ by switching off the interaction between $A$ and $\bar{A}$. 
This leads to the appearance of gapless edge states at the 1D boundary of each piece. 
We can then study the ES of the original system by analyzing how the 1D edge states are entangled as we recover the interaction between $A$ and $\bar{A}$. 

In parallel with this development, there have been active studies on the entanglement between two systems coupled in parallel
such as ladders \cite{Lundgren13, Poilblanc10, Peschel11, Lauchli12, Schliemann12, Furukawa11, Chen13, Tanaka12, Lundgren12, Santos16, Fujita18}, 
bilayers \cite{Schliemann11,Schliemann13,Schliemann14}, an electron-phonon system \cite{Roosz20}, and $d$-dimensional two-component field theories \cite{Xu11,Mollabashi14,Mozaffar16}. 
In a pioneering work, Poilblanc \cite{Poilblanc10} has numerically calculated the ES in gapped phases of a spin-$\frac12$ Heisenberg ladder, 
where the entanglement cut is placed {\it between} the chains, 
and found that the ES remarkably resembles the gapless energy spectrum 
of a single Heisenberg chain (see also \cite{Peschel11,Lauchli12,Schliemann12} for related results). 
This chain-entanglement correspondence is an interesting analogue of the edge-entanglement correspondence in topological phases, 
and these are intimately related through the ``cut and glue'' picture of Qi {\it et al.}\ \cite{Qi12}. 
Field theoretical analyses of the ladder problem have been conducted by describing the system 
as two coupled Tomonaga-Luttinger liquids (TLL) \cite{Lundgren13, Furukawa11, Chen13}. 
It has been shown that the ES exhibits a variety of dispersion relations depending on the type of the interchain coupling. 
Specifically, when the two TLLs (each described by a scalar field) are coupled only by marginal interactions, 
both the symmetric and antisymmetric channels of the scalar fields remain gapless; 
in this fully gapless case, the ES shows a gapped spectrum \cite{Lundgren13,Chen13}. 
When one of the two channels becomes gapped owing to a relevant interaction (i.e., the partially gapless case), 
the ES shows an anomalous square-root dispersion relation \cite{Lundgren13}. 
Finally, when both channels are gapped, the ES shows a linear dispersion relation \cite{Qi12,Lundgren13}, which is consistent with the numerical result by Poilblanc \cite{Poilblanc10}. 
In the fully and partially gapless cases, the ES is thus qualitatively different from the energy spectrum of a single TLL with a linear dispersion relation. 
It has been argued that the unique features of the ES in the gapless cases are associated with 
the emergence of certain long-range interactions in the entanglement Hamiltonian \cite{Lundgren13}. 

In this paper, we consider a system of binary (pseudospin-$\frac12$) Bose-Einstein condensates (BEC) in $d$ spatial dimensions, and study the entanglement between the two components. 
This setup allows us to naturally extend the results of the ladder models in Refs.\ \cite{Lundgren13, Furukawa11, Chen13} to more general $d$-dimensional systems. 
By means of effective field theory, we show that the intercomponent ES exhibits an anomalous square-root dispersion relation 
in the presence of an intercomponent tunneling (a Rabi coupling $\hbar\Omega$) and a gapped dispersion relation in its absence. 
We relate these features of the ES to the emergence of long-range interactions in terms of the superfluid velocity and the particle density in the entanglement Hamiltonian. 
We thus demonstrate that unusual long-range interactions can be emulated in a subsystem of multicomponent BECs that have only short-range interactions. 
We discuss how the emergent long-range interactions are related to the properties of intracomponent correlation functions. 
We also find additive logarithmic contributions to the intercomponent EE that originate from both the Nambu-Goldstone mode 
and the restoration of the global U(1) symmetry for a finite volume in the presence of the Rabi coupling $\hbar\Omega$. 
As we discuss later, our findings have close similarities with the behavior of 
the subregion EE \cite{Song11,Kallin11,Metlitski11,Laflorencie15,Luitz15ssb}, the subregion ES \cite{Metlitski11,Alba13,Kolley13,Rademaker15}, 
and the participation (Shannon-R\'enyi) entropy \cite{Luitz14,Luitz17,Misguich17} 
in systems with spontaneously broken continuous symmetry. 

The rest of this paper is organized as follows. 
In Sec.\ \ref{sec:model}, we introduce the model that we study in this paper, 
and formulate the low-energy effective field theory in terms of the density and phase variables. 
The ground state is then obtained as the vacuum of the Bogoliubov excitations. 
In Sec.\ \ref{sec:RDM}, we derive the expression of the reduced density matrix $\rho_\ua$ for the spin-$\ua$ component by introducing a Gaussian ansatz. 
In Sec.\ \ref{sec:ES}, we calculate the intercomponent ES and the entanglement Hamiltonian in the long-wavelength limit. 
We also analyze the scaling behavior of the intercomponent EE. 
Results are qualitatively different between the cases of $\Omega>0$ and $\Omega=0$. 
In Sec.\ \ref{sec:corr}, we calculate some intracomponent correlation functions 
and discuss their connections with the emergent long-range interactions in the entanglement Hamiltonian. 
In Sec.\ \ref{sec:summary}, we present a summary and an outlook for future studies. 
In Appendices \ref{App:derivation_F_d} and \ref{App:EE}, we describe some technical details of Sec.\ \ref{sec:ES}. 
In Appendix \ref{app:symbr}, we discuss the effect of a weak symmetry-breaking field on entanglement properties. 

\section{Model and effective field theory}\label{sec:model}

We consider a system of binary BECs in $d$ spatial dimensions, 
which has extensively been studied in the context of ultracold atomic gases \cite{Pethick_Smith_2008,Malomed08,Hall08}. 
The Lagrangian density of the system is given by 
\begin{equation}\label{eq:2compBECsLag}
\begin{split}
 \Lcal=& \sum_{\alpha=\ua,\da} 
  \left[ \frac{\irm\hbar}{2} \left( \psi^\dagger_\alpha \dot{\psi}_\alpha - \dot{\psi}^\dagger_\alpha \psi_\alpha \right)
 - \frac{|-\irm\hbar \nabla \psi_\alpha |^2 }{2M} \right]\\
 &- \sum_{\alpha,\beta=\ua,\da} \frac{g_{\alpha\beta}}{2} 
   |\psi_\alpha|^2 |\psi_\beta|^2 
 +\frac{\hbar\Omega}{2}( \psi_\ua^\dagger \psi_\da + \psi_\ua \psi_\da^\dagger ) ,
\end{split}
\end{equation} 
where $\psi_\alpha(\rv,t)$ is the bosonic field for the spin-$\alpha$ component, $M$ is the atomic mass, and $\Omega\ge0$ is the Rabi frequency.  
We assume that the system is confined in a box of volume $V=L^d$ with a periodic boundary condition in every direction. 

We assume contact interactions between atoms. 
For three spatial dimensions, the interaction parameters are given by $g_{\alpha\alpha}=4\pi\hbar^2 a_\alpha/M$ and $g_{\ua\da}=g_{\da\ua}=4\pi\hbar^2 a_{\ua\da}/M$, 
where $a_\alpha$ and $a_{\ua\da}$ are $s$-wave scattering lengths between like and unlike bosons, respectively.
For simplicity, we set $g_{\ua\ua}=g_{\da\da}\equiv g >0$ and $ |g_{\ua\da}| < g$ in the following; 
these conditions ensure the stability of the binary BECs \cite{Pethick_Smith_2008,Malomed08,Ho96,Ao98,Timmermans98,Cazalilla03}. 
While the total number of atoms, $N$, is conserved in this system, the numbers of $\ua$ and $\da$ atoms, $N_\ua$ and $N_\da$, fluctuate in the presence of the Rabi coupling $\Omega>0$. 
We introduce $n=N/(2V)=(N_\ua+N_\da)/(2V)$, which is the average density of $\ua$ and $\da$ atoms. 
We further assume that $N$ is even. 
For $\Omega=0$, where both $N_\ua$ and $N_\da$ are conserved, we assume $N_\ua=N_\da$. 

To describe the low-energy properties of the binary BECs, 
it is useful to decompose the field as $\psi_\alpha = \erm^{-\irm \theta_\alpha} \sqrt{n_\alpha}$, 
where $n_\alpha(\rv,t)$ and $\theta_\alpha(\rv,t)$ are the density and phase variables, respectively.
This type of decomposition has successfully been used to describe scalar BECs \cite{Popov72,Liu98,Khawaja02,Mora03}. 
The validity of our formulation for binary BECs in $d$ dimensions presented in the following is confirmed later 
by the agreement of the excitation spectrum \eqref{eq:E_k} 
with previous results based on other approaches \cite{Goldstein97,Search01,Tommasini03}. 
Furthermore, our formulation has some similarities with those for 1D binary Bose gases in Refs.\ \cite{Whitlock03,Tononi20}.  
The Lagrangian density~\eqref{eq:2compBECsLag} is rewritten in terms of the density and phase variables as 
\begin{equation}\label{eq:Lag_theta_n_pre}
\begin{split}
 \Lcal=&\sum_{\alpha}\left\{\hbar n_\alpha \dot{\theta}_\alpha-\frac{\hbar^2}{2M} 
 \left[n_\alpha(\nabla\theta_\alpha)^2+\frac{(\nabla n_\alpha)^2}{4n_\alpha}\right] \right\} \\
 &-\sum_{\alpha,\beta} \frac{g_{\alpha\beta}}{2} n_\alpha n_\beta 
 +\hbar\Omega \sqrt{n_\ua n_\da}\cos(\theta_\ua-\theta_\da) . 
\end{split}
\end{equation}
In the presence of the Rabi coupling $\Omega>0$,
we can assume that the relative phase $\theta_\ua-\theta_\da$ is locked on average \cite{Leggett91,Tasaki20} 
(i.e., $\langle \theta_\ua-\theta_\da \rangle =0 $)
and its fluctuations acquire a finite mass gap \cite{Goldstein97,Search01,Tommasini03,Whitlock03,Tononi20}. 
To describe this situation, it is useful to make the approximation 
$\cos(\theta_\ua-\theta_\da) \approx 1-(\theta_\ua-\theta_\da)^2/2$ for the last term in Eq.~\eqref{eq:Lag_theta_n_pre} \cite{Lundgren13,Whitlock03,Foini15,Foini17,Tononi20}. 
Since $|n_\alpha(\rv,t) - n| \ll n$ in weakly interacting BECs, we may approximate the Lagrangian density \eqref{eq:Lag_theta_n_pre} 
by taking terms up to the quadratic order in $n_\alpha-n$, $\nabla \theta_\alpha$, and $\theta_\ua-\theta_\da$, obtaining
\begin{equation}\label{eq:Lag_theta_n}
\begin{split}
 \Lcal=&\sum_{\alpha}\left\{\hbar n_\alpha \dot{\theta}_\alpha-\frac{\hbar^2}{2M} 
 \left[n(\nabla \theta_\alpha)^2+\frac{(\nabla n_\alpha)^2}{4n}\right] \right\} \\
 &-\sum_{\alpha,\beta} \frac{g_{\alpha\beta}}{2} n_\alpha n_\beta  \\
 &+\hbar\Omega\left[ \frac{n_\ua+n_\da}{2} -\frac{n}2 (\theta_\ua-\theta_\da)^2-\frac{(n_\ua-n_\da)^2}{8n}\right].
\end{split}
\end{equation}
Henceforth, we ignore the term $\hbar\Omega(n_\ua+n_\da)/2$, which gives only a constant after the spatial integration. 
We note that the resulting expression can also be used for the case of $\Omega=0$ since the approximated part then vanishes. 
Using the Lagrangian \eqref{eq:Lag_theta_n}, the canonical momentum conjugate to $\theta_\alpha$ is found to be 
$\hbar n_\alpha$. We therefore obtain the Hamiltonian density $\mathcal{H}=\sum_\alpha \hbar n_\alpha\dot{\theta}_\alpha - \mathcal{L}$ as 
\begin{equation}\label{eq:Hdensity}
\begin{split}
 \mathcal{H}=&\sum_{\alpha}\frac{\hbar^2}{2M} 
 \left[ n(\nabla\theta_\alpha)^2+\frac{(\nabla n_\alpha)^2}{4n} \right]
 +\sum_{\alpha,\beta}\frac{g_{\alpha\beta}}{2} n_\alpha n_\beta \\
 &+\frac{n\hbar\Omega}{2} (\theta_\ua-\theta_\da)^2 + \frac{\hbar\Omega}{8n} (n_\ua-n_\da)^2. 
\end{split}
\end{equation}


We now quantize the Hamiltonian by requiring the equal-time canonical commutation relation 
$ \left[\theta_\alpha(\rv),n_\beta(\rv')\right]=\irm \delta_{\alpha\beta}\delta(\rv-\rv') $. 
Substituting the Fourier expansions \footnote{
Owing to the compactness $\theta_\alpha\equiv \theta_\alpha+ 2\pi$, 
a winding term $2\pi \bm{M}\cdot\rv/L$ with $\bm{M}\in\mathbb{Z}^d$ is also allowed in the phase $\theta_\alpha(\rv)$, 
as is often discussed in the TLL description \cite{Cazalilla04}. 
Here, we do not include such a term as it vanishes in the ground state. 
}
\begin{equation}\label{eq:FourierTrnf}
 \theta_\alpha(\rv)=\frac{1}{\sqrt{V}}\sum_\kv \theta_{\kv,\alpha}\erm^{\irm\kv\cdot\rv},~~
        n_\alpha(\rv)=\frac{1}{\sqrt{V}}\sum_\kv        n_{\kv,\alpha}\erm^{\irm\kv\cdot\rv}
\end{equation}
in Eq.\ \eqref{eq:Hdensity} and integrating the result over space, 
we obtain the Hamiltonian $H=\int \drm\rv~\mathcal{H}$ as 
\begin{equation}\label{eq:Ham2compBECwavenum}
\begin{split}
 H=\sum_\kv
 \Biggl[ &\sum_{\alpha}\epsilon_\kv 
 \left(n\theta_{-\kv,\alpha}\theta_{\kv,\alpha}+\frac{n_{-\kv,\alpha}n_{\kv,\alpha}}{4n} \right)\\
 &+\sum_{\alpha,\beta}\frac{g_{\alpha\beta}}{2}n_{-\kv,\alpha}n_{\kv,\beta} \\
 &+\frac{n \hbar\Omega}{2} (\theta_{-\kv,\ua}-\theta_{-\kv,\da})(\theta_{ \kv,\ua}-\theta_{ \kv,\da}) \\
 &+\frac{\hbar\Omega}{8n} (n_{-\kv,\ua}-n_{-\kv,\da})(n_{ \kv,\ua}-n_{ \kv,\da}) \Biggr], 
\end{split}
\end{equation}
where $\epsilon_\kv := \hbar^2 \kv^2/(2M)$ is the dispersion relation of a single atom. 
Here, the Fourier components $\theta_{\kv,\alpha} $ and $n_{\kv,\alpha}$ satisfy 
\begin{equation}
\begin{split}
 &[\theta_{\kv,\alpha},n_{-\kv',\beta}]=\irm \delta_{\alpha\beta}\delta_{\kv\kv'}, \\
 &\theta^\dagger_{\kv,\alpha}=\theta_{-\kv,\alpha},~~
        n^\dagger_{\kv,\alpha}=       n_{-\kv,\alpha} ~~ (\alpha,\beta=\ua,\da) .
\end{split}
\end{equation}
The Hamiltonian \eqref{eq:Ham2compBECwavenum} can naturally be decomposed 
in terms of the symmetric and antisymmetric components of the fields. 
Specifically, by introducing  
\begin{equation}\label{eq:theta_n_pm}
  \theta_{\kv,\pm}:=\theta_{\kv,\ua}\pm\theta_{\kv,\da},~~
         n_{\kv,\pm}:=\frac12 (n_{\kv,\ua} \pm n_{\kv,\da}) ,
\end{equation}
which satisfy the commutation relation $[\theta_{\kv,\nu},n_{-\kv',\nu'}]=\irm\delta_{\nu,\nu'}\delta_{\kv,\kv'} ~(\nu,\nu'=\pm)$, 
the Hamiltonian is rewritten as
\begin{equation}\label{eq:H_theta_n_k}
\begin{split}
 H=\sum_{\kv}\sum_{\nu=\pm}
   \bigg[&\frac{n}{2} (\epsilon_\kv+\hbar\Omega\delta_{\nu,-})\theta_{-\kv,\nu}\theta_{\kv,\nu}\\
  &+\frac1{2n} (\epsilon_\kv+2g_{\nu}n) n_{-\kv,\nu}n_{\kv,\nu} \bigg] ,
\end{split}
\end{equation}
where
\begin{equation}
 g_\nu:=g +\nu g_{\ua\da}+\frac{\hbar\Omega}{2n} \delta_{\nu,-}~~ (\nu=\pm).
\end{equation}

We note that for $d=1$, the Hamiltonian \eqref{eq:H_theta_n_k} 
is equivalent, at low energies, to the coupled TLL Hamiltonian studied by Lundgren {\it et al.}\ \cite{Lundgren13} (see Eq.\ (47) and (49) therein)
\begin{equation}\label{eq:H_2TLL}
\begin{split}
 H=\int \drm x \bigg\{ &\sum_{\nu=\pm} \frac{\hbar v_\nu}{2}\left[ K_\nu (\partial_x \vartheta_\nu)^2 + \frac1{K_\nu} (\partial_x\phi_\nu)^2 \right] \\
 &+ \frac{\hbar v_+ m_+^2}{2K_+}\phi_+^2 + \frac{\hbar v_-K_-m_-^2}{2}\vartheta_-^2 \bigg\}
\end{split}
\end{equation}
through the correspondence
\begin{equation}\label{eq:H_2TLL_corres}
\begin{split}
 &\vartheta_\nu=-\frac{\theta_\nu}{\sqrt{2\pi}},~
 \partial_x\phi_\nu = \sqrt{2\pi} \left(n_\nu-n\delta_{\nu,+}\right),\\
 &v_\nu = \sqrt{\frac{g_\nu n}{M}} ,~
 K_\nu=\pi\hbar \sqrt{\frac{n}{g_\nu M}}~~ (\nu=\pm),\\
 &m_+^2=0,~~m_-^2 = \frac{2M\Omega}{\hbar} .
\end{split}
\end{equation}
Here, $v_\pm$ are the sound velocities, $K_\pm$ are the TLL parameters, and $m_\pm^{-1}$ are the correlation lengths in the symmetric ($+$) and antisymmetric ($-$) channels.



In analyzing the Hamiltonian \eqref{eq:H_theta_n_k}, it is convenient to separately treat the part $H^\mathrm{zero}$ corresponding to the zero mode $(\kv=\zerov)$ 
and the part $H^\osc$ corresponding to the oscillator mode $(\kv\neq\zerov)$ 
(see Refs.\ \cite{Lundgren13,Cano15,Metlitski11,Misguich17} for analogous treatments in related contexts).  
Let us first consider the oscillator part $H^\osc$. 
By introducing the annihilation and creation operators as 
\begin{equation}\label{eq:gamma_k}
\begin{split}
 \gamma_{\kv,\nu}=&\frac{1}{\sqrt{2}} 
   \left(\! \sqrt{n}\zeta_{\kv,\nu}\theta_{ \kv,\nu}
   +\frac{\irm n_{ \kv,\nu}}{\sqrt{n}\zeta_{\kv,\nu}}\right),\\
 \gamma_{\kv,\nu}^\dagger=&\frac{1}{\sqrt{2}} 
   \left(\! \sqrt{n}\zeta_{\kv,\nu}\theta_{-\kv,\nu} 
   - \frac{\irm n_{-\kv,\nu}}{\sqrt{n}\zeta_{\kv,\nu}}\right)~~
 (\kv\neq\zerov; \nu=\pm) 
\end{split}
\end{equation}
with 
\begin{equation}\label{eq:zeta_pm}
 \zeta_{\kv,\nu}
 :=\left(\frac{\epsilon_\kv+\hbar\Omega\delta_{\nu,-}}{\epsilon_\kv+2g_{\nu}n}\right)^{1/4}, 
\end{equation}
$H^{\mathrm{osc}}$ is diagonalized as 
\begin{equation}\label{eq:decouple Hamiltonian}
 H^{\mathrm{osc}}=\sum_{\kv\neq\zerov}\sum_{\nu=\pm} 
 E_\nu(\kv)\left(\gamma_{\kv,\nu}^\dag \gamma_{\kv,\nu}+\frac12\right) ,
\end{equation}
where
\begin{equation}\label{eq:E_k}
 E_\nu(\kv)
 :=\sqrt{\left(\epsilon_\kv+\hbar\Omega\delta_{\nu,-}\right)
   \left(\epsilon_\kv+2g_{\nu}n\right)} .
\end{equation}
The excitation spectrum $E_\nu(\kv)$ obtained here reproduces the results of Refs.\ \cite{Goldstein97,Search01,Tommasini03,Whitlock03,Tononi20}. 
For $\Omega=0$, both the symmetric and antisymmetric channels exhibit gapless linear dispersion relations $E_\pm(\kv)\approx (g_\pm n/M)^{1/2}\hbar k$ at low energies; 
for $\Omega>0$, a finite energy gap $\sqrt{2g_-n\hbar\Omega}$ opens in the antisymmetric channel. 
The ground state $\ket{0^\osc}$ of $H^{\osc}$ is specified by the condition that $\gamma_{\kv,\pm}\ket{0^\osc}=0$ for all $\kv\neq\zerov$. 

We next consider the zero-mode part of the Hamiltonian 
\begin{equation}\label{eq:HamZero2compBEC}
 H^{\mathrm{zero}} 
 =\frac{n\hbar\Omega}{2}\theta_{ \zerov,-}^2+\sum_{\nu=\pm} g_\nu n_{\zerov,\nu}^2 ,
\end{equation}
where $n_{\zerov,\pm}$ is related to the atom numbers as $n_{\zerov,\pm}=(N_\ua \pm N_\da)/(2\sqrt{V})$. 
For $\Omega>0$, the $\nu=-$ part of this Hamiltonian can be diagonalized by introducing annihilation and creation operators, 
$\gamma_{\zerov,-}$ and $\gamma_{\zerov,-}^\dagger$ in the same way as Eq.\ \eqref{eq:gamma_k}. Therefore, $H^\zero$ is rewritten as
\begin{equation}\label{eq:Hzero_spec}
 H^{\zero}
 =\frac{g_+}{4V}N^2 
 +E_{\zerov,-}\left(\gamma^\dagger_{\zerov,-} \gamma_{\zerov,-}+\frac12\right) . 
\end{equation}
We consider the ground state of this Hamiltonian for fixed $N$. By analogy with a harmonic oscillator, the ground state in the $n_{\zerov,-}$ basis is given by a gaussian 
\begin{equation}\label{eq:pi0-_GS}
 \bracket{n_{\zerov,-}}{0^\zero}
 \propto \exp\left(-\frac{n_{\zerov,-}^2}{2n\zeta_{\zerov,-}^2}\right) 
  = \exp\left(-\frac{\delta N^2}{N\zeta_{\zerov,-}^2}\right),
\end{equation} 
where $\delta N=(N_\ua - N_\da)/2=\sqrt{V} n_{\zerov,-}$. In terms of the $(N_\ua,N_\da)$ basis, this ground state is expressed as
\begin{equation}\label{eq:GS_Nbasis}
\begin{split}
 \ket{0^\zero} 
 =&\sum_{\delta N}\frac{1}{\sqrt{z}}\exp\left(-\frac{\delta N^2}{N\zeta_{\zerov,-}^2} \right)\\
 &\times\left| N_\ua=\frac{N}{2}+\delta N \right\rangle \left| N_\da=\frac{N}{2}- \delta N \right\rangle ,
\end{split}
\end{equation} 
where $z:=\sum_{\delta N} \exp\left( -\frac{ 2\delta N^2 }{N \zeta_{\zerov,-}^2} \right)$  is the normalization factor. 
For $\Omega=0$, in contrast, $H^\zero$ is given by 
\begin{equation}\label{eq:Hzero_Omega=0}
 H^{\zero}=\frac{g_+}{4V}N^2+\frac{g_-}{4V}(N_\ua-N_\da)^2 .
\end{equation}
In this case, both $N_\ua$ and $N_\da$ are conserved and 
thus the ground state of the Hamiltonian \eqref{eq:Hzero_Omega=0} for the zero mode is given by 
\begin{equation}\label{eq:Omega=0_zero_mode}
  \ket{0^\zero}=\left|N_\ua=N/2\right\rangle \left|N_\da=N/2\right\rangle .
\end{equation}

We note that the effective field theory approach presented here is useful for describing finite-volume BECs 
in which the global symmetry due to particle-number conservation is restored. 
In the mean-field approach, in contrast, the U$(1)$ or U$(1)\times$U$(1)$ symmetry of the original system is completely broken 
in the ground state for $\Omega>0$ and $\Omega=0$, respectively. 
In the latter approach, there is no intercomponent entanglement in the zero-mode ground state. 
Therefore,  the entangled ground state as obtained in Eq.\ \eqref{eq:GS_Nbasis} for $\Omega>0$ 
is a consequence of the global U$(1)$ symmetry restoration. 

For later use, we calculate some correlators in the oscillator-mode part of the spin-$\ua$ component. 
From Eqs.\ \eqref{eq:theta_n_pm} and \eqref{eq:gamma_k}, we find 
\begin{equation}\label{eq:theta_n_gamma}
\begin{split}
  \theta_{\kv,\ua} &= \frac{1}{2\sqrt{2n}} 
  \sum_{\nu=\pm} \zeta_{\kv,\nu}^{-1}\left(\gamma_{\kv,\nu}+\gamma_{-\kv,\nu}^\dagger\right),\\
  n_{\kv,\ua} &=\frac{1}{\irm} \sqrt{\frac{n}{2}} 
  \sum_{\nu=\pm} \zeta_{\kv,\nu}\left(\gamma_{\kv,\nu} -\gamma_{-\kv,\nu}^\dagger\right) .
\end{split}
\end{equation}
The correlators of these operators in the ground state $\ket{0^\osc}$ of $H^\osc$ are then calculated as 
\begin{equation}\label{eq:Correlator_HomoBEC}
\begin{split}
 \bra{0^\osc}\theta_{-\kv,\ua}\theta_{ \kv,\ua}\ket{0^\osc}
 &=\frac{1}{8n} \sum_{\nu=\pm} \zeta_{\kv,\nu}^{-2},\\
 \bra{0^\osc}n_{-\kv,\ua} n_{\kv,\ua}\ket{0^\osc}
  &= \frac{n}{2}\sum_{\nu=\pm}\zeta_{\kv,\nu}^{2}~~~
 (\kv\neq\zerov). 
\end{split}
\end{equation}

\section{Reduced density matrix}\label{sec:RDM}

We now consider the reduced density matrix $\rho_\ua$ for the spin-$\ua$ component,  
which is defined by starting from the ground state $\ket{0^\zero}\otimes\ket{0^\osc}$ of the total system and tracing out the degrees of freedom in the spin-$\da$ component. 
Because the zero and oscillator modes are decoupled, 
the reduced density matrix takes the form $\rho_\ua = \rho_\ua^{\mathrm{zero}} \otimes \rho_\ua^{\mathrm{osc}}$. 
The calculation of the zero-mode part $\rho_\ua^{\mathrm{zero}}$ is rather straightforward as we explain later in Sec.\ \ref{sec:ES}. 
Here we calculate the oscillator-mode part $\rho_\ua^{\mathrm{osc}}$ and the associated ES.

For $\rho_\ua^\osc$, we introduce the following Gaussian ansatz \cite{Lundgren13,Chen13,Metlitski11,Peschel03,Peschel09}: 
\begin{equation}\label{eq:Entanglement_Ham_oscillation}
\begin{split}
 &\rho^{\osc}_\ua =\frac1{Z^{\osc}_\erm} \erm^{-H_\erm^\osc} , ~~Z^\osc_\erm=\Tr~\erm^{-H^{\osc}_\erm} ,\\
 &H^{\mathrm{osc}}_\erm
 =\frac12\sum_{\kv\ne\zerov}
 \left(nF_\kv\theta_{-\kv,\ua}\theta_{\kv,\ua}+\frac{G_\kv}{n}n_{-\kv,\ua}n_{\kv,\ua}\right) ,
\end{split}
\end{equation} 
where $F_\kv$ and $G_\kv$ are positive dimensionless coefficients to be determined later. 
Here, we assume $F_\kv=F_{-\kv}$ and $G_\kv=G_{-\kv}$ without loss of generality. 
By introducing annihilation and creation operators as 
\begin{equation}\label{eq:eta_k}
\begin{split}
  &\eta_\kv=\frac{1}{\sqrt{2}} \left[
    \sqrt{n}             \left(\frac{F_\kv}{G_\kv}\right)^{1/4} \theta_{\kv,\ua} +
    \frac{\irm}{\sqrt{n}}\left(\frac{G_\kv}{F_\kv}\right)^{1/4} n_{\kv,\ua} \right], \\
  &\eta_\kv^\dagger=\frac{1}{\sqrt{2}} \left[
    \sqrt{n}             \left(\frac{F_\kv}{G_\kv}\right)^{1/4} \theta_{-\kv,\ua} -
    \frac{\irm}{\sqrt{n}}\left(\frac{G_\kv}{F_\kv}\right)^{1/4} n_{-\kv,\ua} \right]\\ 
  &(\kv\ne\zerov), 
\end{split}
\end{equation}
we can diagonalize the entanglement Hamiltonian $H_\erm^\osc$ in Eq.\ \eqref{eq:Entanglement_Ham_oscillation} as   
\begin{equation}\label{eq:DiagonalHam}
 H^{\osc}_\erm=\sum_{\kv\ne\zerov} \xi_\kv \left(\eta_\kv^\dagger\eta_\kv+\frac12\right) ,
\end{equation}
where $\xi_\kv:=\sqrt{F_\kv G_\kv}$ is the single-particle ES. 

Using the relations in Eq. \eqref{eq:eta_k} and the Bose distribution function 
\begin{equation}\label{eq:fB_xi}
 \Tr\left(\eta_\kv^\dagger \eta_\kv \rho^\osc_\ua\right)=\frac{1}{\erm^{\xi_\kv}-1} =: f_\Brm(\xi_\kv) ,
\end{equation}
we obtain phase and density correlators as  
\begin{subequations}\label{eq:Correlator_FkGk} 
\begin{align}
 &\Tr\left(\theta_{-\kv,\ua}\theta_{ \kv,\ua}\rho^\osc_\ua\right) \notag\\
 &=\frac{1}{2n}\left(\frac{G_\kv}{F_\kv}\right)^{1/2} \!\!
   \Tr\left[\left(\eta_{ \kv}^\dagger+\eta_{-\kv}\right) 
              \left(\eta_{ \kv}+\eta_{-\kv}^\dagger\right)\rho^\osc_\ua\right] \notag\\
 &=\frac{1}{n} \left(\frac{G_\kv}{F_\kv}\right)^{1/2} \! \left[f_\Brm(\xi_\kv)+\frac12\right],\\
 &\Tr\left(n_{-\kv,\ua}n_{\kv,\ua}\rho^\osc_\ua\right) \notag\\
 &=\frac{n}{2}\left(\frac{F_\kv}{G_\kv}\right)^{1/2} \!\!
     \Tr\left[\left(\eta_{ \kv}^\dagger-\eta_{-\kv} \right)
                 \left(\eta_{ \kv} -\eta_{-\kv}^\dagger\right) 
                \rho^\osc_\ua\right]  \notag\\
 &=n\left(\frac{F_\kv}{G_\kv}\right)^{1/2} \! \left[f_\Brm(\xi_\kv)+\frac12\right] . 
\end{align} 
\end{subequations}
By requiring these to be equal to the correlators \eqref{eq:Correlator_HomoBEC} 
calculated for the oscillator-mode ground state $\ket{0^\mathrm{osc}}$, we obtain
\begin{align}
  f_\Brm(\xi_\kv)
  &=\sqrt{\bra{0^\osc}\theta_{-\kv,\ua}\theta_{\kv,\ua}\ket{0^\osc}
     \bra{0^\osc}n_{-\kv,\ua}n_{\kv,\ua}\ket{0^\osc}} - \frac12  \notag\\
  &=\frac{\zeta_{\kv,+}^2+\zeta_{\kv,-}^2}{4\zeta_{\kv,+}\zeta_{\kv,-}}-\frac12,
  \label{eq:BoseDistributionFunc_correlator} \\
  \sqrt{\frac{F_\kv}{G_\kv}} 
  &=\frac{1}{n}\sqrt{\frac{\bra{0^\osc}n_{-\kv,\ua}n_{\kv,\ua}\ket{0^\osc}}
     {\bra{0^\osc}\theta_{-\kv,\ua}\theta_{\kv,\ua}\ket{0^\osc}} }
  =2\zeta_{\kv,+}\zeta_{\kv,- } ,
  \label{eq:EntanglementHam_s_k}
\end{align}
from which we obtain the single-particle ES $\xi_\kv$ and the coefficients $F_\kv$ and $G_\kv$ as
\begin{subequations}\label{eq:xi_F_G_general}
\begin{align}
  &\xi_\kv=\ln\left[1+\frac{1}{f_\Brm(\xi_\kv)}\right] =2\ln \frac{\zeta_{\kv,+}+\zeta_{\kv,-}}{|\zeta_{\kv,+}-\zeta_{\kv,-}|}, \label{eq:xi_general}\\
  &F_\kv=2\xi_\kv \zeta_{\kv,+}\zeta_{\kv,- } ,~~ 
  G_\kv=\frac{\xi_\kv}{2\zeta_{\kv,+}\zeta_{\kv,-}} .
\end{align}
\end{subequations}

\section{Entanglement properties}\label{sec:ES}

In this section, we perform the long-wavelength (i.e., small-$k$) expansion of Eq.\ \eqref{eq:xi_F_G_general}, 
and discuss the properties of the ES and the entanglement Hamiltonian. 
Using the obtained ES, we also analyze the scaling behavior of the intercomponent EE. 
Results are qualitatively different between the cases of $\Omega>0$ and $\Omega=0$, 
which we discuss separately in Secs.\ \ref{sec:ES_Omega>0} and \ref{sec:ES_Omega=0}. 
We introduce scaled coupling constants
\begin{equation}
\begin{split}
 \gt_\nu &:= \frac{2M}{\hbar^2} \times 2g_\nu n=\frac{4g_\nu nM}{\hbar^2}~(\nu=\pm),\\
 \Omegat &:=\frac{2M}{\hbar^2}\times \hbar\Omega=\frac{2M \Omega}{\hbar} 
\end{split}
\end{equation}
so that $\zeta_{\kv,\pm}$ defined in Eq.\ \eqref{eq:zeta_pm} can be expressed simply as 
\begin{equation}\label{eq:zeta_pm_simp}
 \zeta_{\kv,\nu} = \left( \frac{\Omegat\delta_{\nu,-}+k^2}{\gt_\nu+k^2} \right)^{1/4}. 
\end{equation}

\subsection{Case of $\Omega>0$}\label{sec:ES_Omega>0}

\subsubsection{Entanglement spectrum and Hamiltonian}

Let us first consider the case of a finite Rabi coupling $\Omega>0$. 
Since $\ket{0^\zero}$ in Eq.\ \eqref{eq:GS_Nbasis} is already written in the form of the Schmidt decomposition, 
the reduced density matrix for the zero mode is obtained as 
\begin{equation}\label{eq:rho_zero_Omega>0}
\begin{split}
 \rho^{\mathrm{zero}}_{\ua} 
 = \sum_{\delta N} &\frac{1}{z} \exp\left( -\frac{2\delta N^2}{N\zeta_{\zerov,-}^2} \right)\\
 &\times \left|N_\ua=\frac{N}{2}+\delta N \right\rangle
 \left\langle N_\ua=\frac{N}{2}+\delta N \right| .
\end{split}
\end{equation}
The associated entanglement Hamiltonian is thus given by
\begin{equation}\label{eq:EntanglementHam_zero}
\begin{split}
 H_\erm^{\mathrm{zero}}
 &= \frac{2 \delta N^2 }{N \zeta_{\zerov,-}^2}  
 = 2\left( \frac{\gt_-}{\Omegat} \right)^{1/2} \frac{ (N_\ua-N/2)^2 }{N} \\
 &= \frac{G_0}{2nV} (N_\ua-N/2)^2, 
\end{split}
\end{equation}
where $G_0=2\left(\gt_-/\Omegat\right)^{1/2}$ as defined in Eq.\ \eqref{eq:FG_Omega>1} below. 

For the oscillator-mode part, we perform the long-wavelength expansion of $\xi_\kv, F_\kv$ and $G_\kv$ in Eq.\ \eqref{eq:xi_F_G_general} 
by assuming $\epsilon_\kv \ll 2g_\pm n, \hbar \Omega$. 
From Eq.\ \eqref{eq:zeta_pm_simp}, we find 
\begin{equation}\label{eq:zeta_pm_Omega>0}
\begin{split}
\zeta_{\kv,+}&=\frac{k^{1/2}}{\gt_+^{1/4}} \left[1+O(k^2)\right],\\
\zeta_{\kv,-}&= \left(\frac{\Omegat}{\gt_-}\right)^{1/4}\left[1+O(k^2)\right].
\end{split}
\end{equation}
The expansion of $\xi_\kv$ in terms of $k$ is then obtained as
\begin{equation}\label{eq:xi_k_Omega>0}
\begin{split}
 \xi_\kv &= 2\ln \frac{1+\zeta_{\kv,+}/\zeta_{\kv,-}}{1-\zeta_{\kv,+}/\zeta_{\kv,-}}
 =4\left(\frac{\zeta_{\kv,+}}{\zeta_{\kv,-}}\right) + \frac43 \left( \frac{\zeta_{\kv,+}}{\zeta_{\kv,-}} \right)^3 +\dots\\
 &=c_{1/2}k^{1/2}+c_{3/2}k^{3/2}+O(k^{5/2})
\end{split}
\end{equation}
with the coefficients
\begin{equation}\label{eq:xi_k_Omega>0_coeff}
 c_{1/2}=4\left(\frac{\gt_-}{\gt_+ \Omegat}\right)^{1/4},~
 c_{3/2}=\frac43 \left(\frac{\gt_-}{\gt_+ \Omegat}\right)^{3/4}.
\end{equation}
We further obtain
\begin{equation}\label{eq:FkGk_Omega>1}
\begin{split}
 F_\kv&=2\xi_\kv\zeta_{\kv,+}\zeta_{\kv,-}=F_1k +F_2k^2 +O\left(k^3\right),\\
 G_\kv&=\frac{\xi_\kv}{2\zeta_{\kv,+}\zeta_{\kv,-}} =G_0 +G_1k +O\left(k^2\right) 
\end{split}
\end{equation}
with the coefficients 
\begin{equation}\label{eq:FG_Omega>1}
\begin{split}
 F_1&=2c_{1/2}\left(\frac{\Omegat}{\gt_+\gt_-}\right)^{1/4}=\frac{8}{\gt_+^{1/2}},\\
 F_2&=2c_{3/2}\left(\frac{\Omegat}{\gt_+\gt_-}\right)^{1/4}=\frac{8\gt_-^{1/2}}{3\gt_+\Omegat^{1/2}},\\
 G_0&=\frac{c_{1/2}}{2}\left(\frac{\gt_+\gt_-}{\Omegat}\right)^{1/4}=\frac{2\gt_-^{1/2}}{\Omegat^{1/2}} ,\\
 G_1&=\frac{c_{3/2}}{2}\left(\frac{\gt_+\gt_-}{\Omegat}\right)^{1/4}=\frac{2\gt_-}{3\gt_+^{1/2}\Omegat}.
\end{split}
\end{equation}
Interestingly, the single-particle ES $\xi_\kv$ is proportional to $\sqrt{k}$ in the long-wavelength limit. 
This anomalous dispersion relation is associated with emergent long-range interactions 
in the entanglement Hamiltonian as we explain in the following. 

The total entanglement Hamiltonian $H_\erm$ in the long-wavelength limit is given by the sum of 
the zero-mode part $H_\erm^{\mathrm{zero}}$ [Eq.\ \eqref{eq:EntanglementHam_zero}] 
and the oscillator-mode part $H_\erm^{\mathrm{osc}}$ [Eq. \eqref{eq:Entanglement_Ham_oscillation} with Eq.\ \eqref{eq:FkGk_Omega>1}].
Using the real-space representation of the fields $\theta_\ua(\rv)$ and $n_\ua(\rv)$, it can be expressed as 
\begin{equation}\label{eq:He_Omega_Real}
\begin{split}
 H_\erm 
 =&\int\mathrm{d}^d\rv \int\mathrm{d}^d\rv'\frac{nF_1}{2} U_d(\rv-\rv') \nabla\theta_\ua(\rv) \cdot \nabla\theta_\ua(\rv') \\
   &+\int\mathrm{d}^d\rv~\frac{G_0}{2n} \left[ n_\ua(\rv)-n\right]^2
 +\int\mathrm{d}^d\rv~\frac{nF_2}{2}\left[ \nabla\theta_\ua(\rv) \right]^2\\
  &+\int\mathrm{d}^d\rv \int\mathrm{d}^d\rv'~\frac{G_1}{2n}U_d(\rv-\rv') \nabla n_\ua(\rv)\cdot\nabla'n_\ua(\rv')  \\
  &+\dots. 
\end{split}
\end{equation}
Here, we introduce the long-range interaction potential 
\begin{equation}\label{eq:F_d_Bare}
\begin{split}
  &U_{d}(\rv-\rv'):=\lim_{\alpha\to0^+} U_d(\rv-\rv';\alpha),\\
  &U_d(\rv-\rv';\alpha) :=\sum_{\kv\neq\zerov} \frac{1}{Vk} \erm^{-\alpha k +\irm\kv\cdot(\rv-\rv')},
\end{split}
\end{equation}
where we use the convergence factor $\erm^{-\alpha k}$ to regularize the infinite sum. 
As described in Appendix \ref{App:derivation_F_d}, this potential is calculated for $d=1,2,3$ as 
\begin{equation}\label{eq:U123}
\begin{split}
 U_{1}(x-x')&= -\frac{1}{\pi}\ln\frac{2\pi D(x-x'|L)}{L} ,\\
 U_{2}(\rv-\rv')&=\frac{1}{2\pi |\rv-\rv'|} , \\
 U_{3}(\rv-\rv')&=\frac{1}{2\pi^2 |\rv-\rv'|^2} .
\end{split}
\end{equation}
Here, we introduce the chord distance \cite{Cazalilla04}
\begin{equation}\label{eq:chord}
 D(x-x'|L)=\frac{L}{\pi} \left| \sin\frac{\pi (x-x')}{L} \right|= \frac{L}{2\pi} \left|\erm^{\irm\frac{2\pi}{L}x}-\erm^{\irm\frac{2\pi}{L}x'}\right|,
\end{equation}
which is the length of the chord between two points separated by an arc length $|x-x'|$ on a ring of circumference $L$; 
for $|x-x'|\ll L$, $D(x-x'|L)\approx |x-x'|$. 
It is interesting to compare Eq.\ \eqref{eq:He_Omega_Real} with the spin-$\ua$ part of the original Hamiltonian \eqref{eq:Hdensity}. 
In addition to local terms, the entanglement Hamiltonian  \eqref{eq:He_Omega_Real} has long-range interactions in terms of 
the superfluid velocity $\vv_{\srm,\ua}(\rv)=-\frac{\hbar}{M}\nabla\theta_\ua(\rv)$ and the density gradient $\nabla n_\ua(\rv)$. 
In particular, the long-range interaction in terms of $ \vv_{\srm,\ua}(\rv)$ is crucial for the emergence of the anomalous dispersion relation $\xi_\kv\propto \sqrt{k}$ in the ES. 
For $d=1$, the same long-range interaction $U_1(x-x')$ has also been found in the entanglement Hamiltonian for coupled TLLs \cite{Lundgren13} 
and the ground-state wave functional of the TLL \cite{Furukawa11,Fradkin93,Stephan09}. 

\subsubsection{Entanglement entropy} 

Using the ES obtained above, we proceed to calculate the intercomponent EE $S_\mathrm{e}$. 
As shown in Appendix \ref{App:EE_gapless} [see Eqs.\ \eqref{eq:Se_osc_gapless} and \eqref{eq:Se_osc_gapless_EM} therein], 
the oscillator-mode part gives the contribution
\begin{equation}\label{eq:Se_osc_Omega>0}
 S_\erm^\osc =  \frac{\sigma L^d}{c_{1/2}^{2d}} -\frac12 \ln \frac{L}{(2\pi) c_{1/2}^2}+O(1).
\end{equation}
Here, the leading contribution is given by the first term, which is proportional to the volume $L^d$ of the system with a non-universal coefficient $\sigma$. 
Such a volume-law contribution is standard for an extensive cut as discussed here, 
and has also been found in other systems \cite{Furukawa11,Chen13,Lundgren13,Xu11,Mollabashi14}. 
Besides, there is a subleading logarithmic contribution with the universal coefficient $-1/2$, 
which is identified through a careful examination of small-$k$ contributions 
as described in Appendix \ref{App:EE_gapless} and therefore originates from the Nambu-Goldstone mode. 

We next calculate the zero-mode contribution $S_\erm^\zero$ to the EE.  
To this end, we consider the canonical ensemble given 
by the zero-mode entanglement Hamiltonian \eqref{eq:EntanglementHam_zero} at a fictitious temperature $T$, 
and calculate the partition function as 
\begin{equation}
 Z_\mathrm{e}^\zero=\sum_{\delta N=-\infty}^\infty \exp\left[ - \frac{G_0}{2nVT} (\delta N)^2 \right] \approx \sqrt{\frac{2\pi nV T}{G_0}},
\end{equation}
where we approximate the sum by a Gaussian integral. 
The contribution to the EE is then calculated as
\begin{equation}\label{eq:Se_zero_Omega>0}
 S_\mathrm{e}^\zero=\frac{\partial}{\partial T} \left(T\ln Z_\erm^\zero\right) \bigg|_{T=1}
 = \frac12 \ln \frac{2\pi \erm nV }{G_0}.
\end{equation}

The total EE is given by the sum of Eqs.\ \eqref{eq:Se_osc_Omega>0} and \eqref{eq:Se_zero_Omega>0}, i.e., 
\begin{equation}\label{eq:Se_Omega>0}
\begin{split}
  S_\erm &= S_\erm^\zero+S_\erm^\osc \\
  &=  \frac{\sigma L^d}{c_{1/2}^{2d}}+\frac{d}{2} \ln \left[ \left(\frac{2\pi \erm n}{G_0}\right)^{1/d} L \right] -\frac12 \ln \frac{L}{(2\pi) c_{1/2}^2}+O(1).
\end{split}
\end{equation}
This expression includes the leading volume-law term from the oscillator mode
as well as the subleading logarithimic terms from both the zero and oscillator modes. 
The coefficient of the logarithmic contribution is $(d-1)/2$ in total. 
The intercomponent EE per volume in the thermodynamics limit (i.e., the EE density) is given by
\begin{equation}\label{eq:Se_Omega_pV}
 \lim_{L\to\infty} \frac{S_\erm}{L^d} = \frac{\sigma}{c_{1/2}^{2d}} = \frac{\sigma}{4^{2d}} \left(\frac{\gt_+\Omegat}{\gt_-} \right)^{d/2},
\end{equation} 
where we use Eq.\ \eqref{eq:xi_k_Omega>0_coeff}.
Aside from the non-universal coefficient $\sigma$, which is expected to depend only weakly on the system's parameters, 
Eq.\ \eqref{eq:Se_Omega_pV} is a monotonically increasing function of $g_{\ua\da}/g$ and $\Omega$. 
Interestingly, an intercomponent attraction $g_{\ua\da}<0$ leads to a reduction in the EE density. 

It is interesting to compare the present results with the behavior of the subregion ES and EE 
in systems with spontaneously broken continuous symmetry. 
We have seen a clear decoupling of the zero- and oscillator-mode contributions to the entanglement Hamiltonian and thus an analogous decoupling in the ES. 
This behavior is similar to those found in the subregion ES \cite{Metlitski11,Alba13,Kolley13,Rademaker15}. 
In particular, a quadratic dependence of the zero-mode entanglement Hamiltonian \eqref{eq:EntanglementHam_zero} on the subsystem particle number, 
which is reminiscent of a ``tower of states'' spectrum, 
has also been found in the subregion ES of the Bose-Hubbard model \cite{Alba13}. 
In the intercomponent EE, we have found a subleading logarithmic contribution with a coefficient $(d-1)/2$. 
Similar logarithmic contributions have also been found in the subregion EE in Refs.\ \cite{Song11,Kallin11,Metlitski11,Laflorencie15,Luitz15ssb} 
(see also Refs.\ \cite{Luitz14,Luitz17,Misguich17} for similar behavior in the participation entropy). 
In particular, Metlitski and Grover \cite{Metlitski11} have shown that such a contribution 
has the universal coefficient $N_\mathrm{NG}(d-1)/2$ with $N_\mathrm{NG}$ being the number of Nambu-Goldstone modes 
when the subregion boundary is smooth (i.e., has no corner). 
Here, the coefficient $N_\mathrm{NG}(d-1)/2$ arises solely from the restoration of symmetry for a finite volume, 
and is related to the fact that the subregion has the boundary size proportional to $R^{d-1}$, where $R$ is the length scale of the subregion; 
the oscillator mode does not give a logarithmic contribution as there is no other length scale as far as the subsystem boundary is smooth. 
In contrast, the coefficient $(d-1)/2$ obtained for the intercomponent EE in this section is contributed from the zero mode and the oscillator mode. 
From Eq.\ \eqref{eq:Se_osc_Omega>0}, we find that the emergence of the logarithmic oscillator-mode contribution is 
related to an additional length scale $c_{1/2}^2=16(\gt_-/\gt_+\Omegat)^{1/2}$ caused by the intercomponent couplings. 

We have discussed the importance of the symmetry restoration in obtaining the logarithmic behavior in Eq.\ \eqref{eq:Se_zero_Omega>0}. 
In Appendix \ref{app:symbr}, we discuss the effect of a weak symmetry-breaking field $\tilde{h}$ on the behavior of the EE. 
As explained there, the behavior of $S_\erm^\zero$ is easily influenced by weak $\tilde{h}$ with $\tilde{h}\gg \gt_+/(64n^2V^2)$, 
resulting in a logarithmic dependence on $\tilde{h}$ [see Eq.\ \eqref{eq:EE_zero_ht}]. 
If we tune $\tilde{h}$ to the cutoff $\gt_+/(64n^2V^2)$ in the final expression, 
we can reproduce the logarithmic behavior in Eq.\ \eqref{eq:Se_zero_Omega>0}. 
This is consistent with the argument in Refs.\ \cite{Song11,Laflorencie15,Luitz15ssb} 
that calculations with a properly chosen symmetry-breaking field can mimic the behavior of the symmetry-restored ground state.  

In passing, we note that our result \eqref{eq:Se_osc_Omega>0} on the oscillator-mode contribution to the EE 
is based on the the leading term in the single-particle ES \eqref{eq:xi_k_Omega>0}, 
which is valid for small $k$ with $\epsilon_\kv \ll 2g_\pm n, \hbar \Omega$. 
As we decrease $\Omega$, the validity range of Eq.\ \eqref{eq:xi_k_Omega>0} is restricted to smaller $k$. 
One then needs a larger size $L$ to clearly see the scaling in Eq.\ \eqref{eq:Se_osc_Omega>0}. 
In the limit $\Omega\to 0$, Eq.\ \eqref{eq:Se_osc_Omega>0} completely loses its validity. 
Therefore, we need a separate discussion for $\Omega=0$, which is given next. 



\subsection{Case of $\Omega=0$}\label{sec:ES_Omega=0}

\subsubsection{Entanglement spectrum and Hamiltonian} 

Let us next consider the case in which the Rabi coupling $\Omega$ is not present. 
In this case, the ground state \eqref{eq:Omega=0_zero_mode} of the Hamiltonian \eqref{eq:Hzero_Omega=0} for the zero mode is a product state, 
and thus gives no contribution to the entanglement Hamiltonian.
For the oscillator-mode part, we perform the long-wavelength expansion of $\xi_\kv, F_\kv$ and $G_\kv$ in Eq.\ \eqref{eq:xi_F_G_general} 
by assuming $\epsilon_\kv \ll 2g_\pm n$. 
From Eq.\ \eqref{eq:zeta_pm_simp}, we find 
\begin{equation}\label{eq:zeta_pm_Omega=0}
 \zeta_{\kv,\nu} = \frac{k^{1/2}}{\gt_\nu^{1/4}} \left(1+\frac{k^2}{\gt_\nu} \right)^{-1/4} 
 = \frac{k^{1/2}}{\gt_\nu^{1/4}} \left[ 1-\frac{k^2}{4\gt_\nu} +O\left(k^4\right) \right]. 
\end{equation}
The expansion of $\xi_\kv$ in terms of $k$ is then obtained as
\begin{equation}\label{eq:xi_k_Omega=0}
\begin{split}
 \xi_\kv &= 2\ln \frac{\zeta_{\kv,+}^{-1}+\zeta_{\kv,-}^{-1}}{|\zeta_{\kv,+}^{-1}-\zeta_{\kv,-}^{-1}|}\\
 &= 2\ln \frac{\gt_+^{1/4}+\gt_-^{1/4}+\frac14 \left(\gt_+^{-3/4}+\gt_-^{-3/4} \right)k^2+O(k^4)}
   {\big|\gt_+^{1/4}-\gt_-^{1/4}+\frac14 \left(\gt_+^{-3/4}-\gt_-^{-3/4} \right)k^2+O(k^4)\big|}\\
 &=\xi_0+c_2 k^2+O(k^4)
\end{split}
\end{equation}
with the coefficients
\begin{equation}
\begin{split}
 \xi_0&=2\ln \frac{\gt_+^{1/4}+\gt_-^{1/4}}{\big|\gt_+^{1/4}-\gt_-^{1/4}\big|},\\
 c_2&=\frac12\left( \frac{\gt_+^{-3/4}+\gt_-^{-3/4} }{\gt_+^{1/4}+\gt_-^{1/4}}-\frac{\gt_+^{-3/4}-\gt_-^{-3/4} }{\gt_+^{1/4}-\gt_-^{1/4}} \right)\\
 &=\frac{\gt_+^{1/2}+\gt_-^{1/2}}{(\gt_+\gt_-)^{3/4}}.
\end{split}
\end{equation}
We further obtain
\begin{equation}\label{eq:Fk_Gk_withoutRabi}
\begin{split}
 F_\kv &= 2\xi_\kv \zeta_{\kv,+}\zeta_{\kv,-} = F_1k +F_3k^3+O(k^5) ,\\
 G_\kv &=\frac{\xi_\kv}{2\zeta_{\kv,+}\zeta_{\kv,-}}=G_{-1}k^{-1}+G_1k+O(k^3) 
\end{split}
\end{equation}
with the coefficients
\begin{equation}\label{eq:F_G_coeff_Omega=0}
\begin{split}
 F_1&=\frac{2\xi_0}{(\gt_+\gt_-)^{1/4}},\\
 F_3&=\frac{2}{(\gt_+\gt_-)^{1/4}} \left[ c_2 - \frac14 \left(\frac1{\gt_+}+\frac1{\gt_-} \right)\xi_0 \right],\\
 G_{-1}&=\frac12 (\gt_+\gt_-)^{1/4} \xi_0,\\
 G_1 &= \frac12 (\gt_+\gt_-)^{1/4} \left[ c_2 + \frac14 \left(\frac1{\gt_+}+\frac1{\gt_-} \right)\xi_0 \right].
\end{split}
\end{equation}
Interestingly, the single-particle ES \eqref{eq:xi_k_Omega=0} is gapped for $\Omega=0$ 
in contrast to the gapless ES \eqref{eq:xi_k_Omega>0} for $\Omega>0$. 

Using Eqs. \eqref{eq:Entanglement_Ham_oscillation} and \eqref{eq:Fk_Gk_withoutRabi} and focusing on the leading contributions, 
we obtain the real-space representation of the entanglement Hamiltonian as
\begin{equation}\label{eq:He_Omega_Real_Omega=0}
\begin{split}
 H_\erm=\int\mathrm{d}^d\rv \int\mathrm{d}^d\rv'&~U_d(\rv-\rv')
 \Biggr[\frac{nF_1}{2} \nabla\theta_\ua(\rv) \cdot \nabla\theta_\ua(\rv') \\
  &+\frac{G_{-1}}{2n} \left[n_\ua(\rv)-n\right] \left[n_\ua(\rv')-n\right] 
  \Biggr] +\dots. 
\end{split}
\end{equation}
We thus find that long-range interactions in terms of the superfluid velocity $\vv_{\srm,\ua}(\rv)=-\frac{\hbar}{M}\nabla\theta_\ua(\rv)$ and the density $n_\ua(\rv)$ emerge in the entanglement Hamiltonian, 
which are crucial for the emergence of the gapped ES.  
In contrast to the case of $\Omega>0$ shown in Eq.\ \eqref{eq:He_Omega_Real}, the entanglement Hamiltonian \eqref{eq:He_Omega_Real_Omega=0} 
contains neither $F_2$ nor $G_0$ present in the original Hamiltonian \eqref{eq:Hdensity}. 

\subsubsection{Entanglement entropy} 

As there is no intercomponent entanglement in the zero mode, 
we focus on the oscillator-mode contribution $S_\erm^\osc$ to the EE. 
As described in Appendix \ref{App:EE_gapped} [see Eqs.\ \eqref{eq:Se_osc_gapped} and \eqref{eq:Se_osc_gapped_coeff} therein], the gapped dispersion relation \eqref{eq:xi_k_Omega=0} in the ES leads to 
a volume-law scaling followed by the negative universal constant: 
\begin{equation}\label{eq:Se_Omega=0}
 S_\mathrm{e}^\osc =s_1 V-s_0.
\end{equation}
Here, $s_1$ depends on the ultraviolet cutoff of the theory and is thus non-universal. 
The negative constant contribution $-s_0$ is due to the lack of entanglement in the zero-mode ground state. 
The constant $s_0$ is a universal function of the coupling ratio $g_-/g_+$, and is given by 
\begin{equation}\label{eq:s0_Omega=0} 
 s_0 =\frac{\xi_0}{\erm^{\xi_0}-1}-\ln \left( 1-\erm^{-\xi_0}\right).
\end{equation}
For $d=1$, this expression is consistent with the universal constant obtained for coupled TLLs \cite{Lundgren13,Furukawa11,Chen13} 
through the correspondence $g_-/g_+=(K_+/K_-)^2$, where $K_\pm$ are the TLL parameters [see Eq.\ \eqref{eq:H_2TLL_corres}]. 


\section{Intracomponent correlation functions}\label{sec:corr}

In this section, we calculate some intracomponent correlation functions, and discuss their connections with the long-wavelength entanglement Hamiltonian $H_\erm$ obtained in Sec.\ \ref{sec:ES}. 
Let $\langle{\cal O}\rangle$ denote the expectation value of an operator ${\cal O}$ with respect to the ground state $\ket{0^\zero}\otimes\ket{0^\osc}$ of the total system. 
If ${\cal O}$ acts only on the spin-$\ua$ component, $\langle{\cal O}\rangle$ should be equal to $\Tr \left({\cal O} \erm^{-H_\erm}\right) / \Tr~\erm^{-H_\erm}$ 
as far as long-distance properties are concerned.
Our purpose here is to investigate how the unusual long-range interactions in $H_\erm$ manifest themselves in the correlation properties of the system. 

\subsection{Case of $\Omega>0$}\label{sec:corr_Omega>0}

Owing to the gapless ES $\xi_\kv$, we can approximate the Bose distribution function \eqref{eq:fB_xi} as 
$f_\Brm(\xi_\kv)\approx \xi_\kv^{-1}=(F_\kv G_\kv)^{-1/2}$ for sufficiently small $k$. 
Then, in the long-wavelength limit, Eq.\ \eqref{eq:Correlator_FkGk} gives
\begin{subequations}\label{eq:corr_theta_n_Omega>0_FG}
\begin{align}
 \langle \theta_{-\kv,\ua}\theta_{ \kv,\ua} \rangle &\approx \frac{1}{nF_\kv}\approx \frac{1}{nF_1k}, \label{eq:corr_theta_n_Omega>0_F}\\
 \langle n_{-\kv,\ua} n_{\kv,\ua} \rangle &\approx  \frac{n}{G_\kv}\approx \frac{n}{G_0}~~(\kv\ne\zerov), \label{eq:corr_theta_n_Omega>0_G}
\end{align}
\end{subequations}
where we use Eq.\ \eqref{eq:FkGk_Omega>1}. 
Therefore, the phase and density fluctuations are directly related to the coefficients $F_\kv$ and $G_\kv$, respectively, in the entanglement Hamiltonian. 
Equation \eqref{eq:corr_theta_n_Omega>0_FG} also indicates that an increase in $F_\kv$ ($G_\kv$) leads to a suppression of the phase (density) fluctuation, 
which can be understood naturally from the expression of $H_\erm^\osc$ in Eq.\ \eqref{eq:Entanglement_Ham_oscillation}. 

We also discuss the $\kv=\zerov$ component of the correlations. 
Using Eq.\ \eqref{eq:EntanglementHam_zero}, we can calculate the variance of the spin-$\ua$ atom number $N_\ua$ around its mean value $N/2$ as
\begin{equation}\label{eq:corr_Nup}
 \left\langle \left( N_\ua-N/2 \right)^2 \right\rangle
 = \frac1z \sum_{\delta N} (\delta N)^2 \exp\left( -\frac{G_0 (\delta N)^2}{2nV} \right) \approx \frac{nV}{G_0},
\end{equation}
where we approximate the sum by a Gaussian integral. 
Using $N_\ua = \sqrt{V} n_{\zerov,\ua}$ and $N/2=nV$, we equivalently have 
\begin{equation}\label{eq:corr_n0up}
 \left\langle \left( n_{\zerov,\ua}-n\sqrt{V} \right)^2 \right\rangle = \frac1V \left\langle \left( N_\ua-N/2 \right)^2 \right\rangle \approx \frac{n}{G_0} ,
\end{equation}
where the connection with the $\kv\ne\zerov$ case in Eq.\ \eqref{eq:corr_theta_n_Omega>0_G} can be seen more clearly. 
Equation \eqref{eq:corr_Nup} indicates that an increase in $G_0$ also leads to a suppression of the fluctuation of $N_\ua$. 
We note that the correlation related to $\theta_{\zerov,\ua}$ cannot be determined as $\theta_{\zerov,+}$ fluctuates completely for fixed $N$. 

We proceed to analyze the one-particle density matrix 
\begin{equation}\label{eq:corr_psi_theta_n}
 \langle \psi_\ua(\rv)^\dagger \psi_\ua(\zerov)\rangle=\left\langle \sqrt{n_\ua(\rv)} \erm^{\irm (\theta_\ua(\rv)-\theta_\ua(\zerov))} \sqrt{n_\ua(\zerov)}\right\rangle,
\end{equation}
which plays a key role in the characterization of the Bose-Einstein condensation \cite{Pethick_Smith_2008, Penrose56,Yang62}. 
Its long-distance behavior is determined dominantly by the phase fluctuation 
as seen in the small-$k$ behavior of Eq.\ \eqref{eq:corr_theta_n_Omega>0_FG}. 
Using Eq.\ \eqref{eq:corr_theta_n_Omega>0_F}, the phase correlation function in real space is obtained as
\begin{equation}
\begin{split}
 \langle \left[ \theta_\ua(\rv)-\theta_\ua(\zerov) \right]^2\rangle 
 &= \frac2V \sum_{\kv\ne\zerov} \erm^{-\alpha k} \left[1- \cos \left({\kv\cdot\rv}\right) \right] \langle \theta_{-\kv,\ua}\theta_{ \kv,\ua} \rangle\\
 &\approx \frac{2}{nF_1} \left[ U_d(\zerov;\alpha)-U_d(\rv;\alpha) \right],
\end{split}
\end{equation}
where we introduce the convergence factor $\erm^{-\alpha k}$ to regularize the infinite sum 
and the function $U_d(\rv;\alpha)$ is defined in Eq.\ \eqref{eq:F_d_Bare} and calculated in Appendix \ref{App:derivation_F_d}. 
The one-particle density matrix is then obtained as 
\begin{equation}\label{eq:corr_psi_Ud}
\begin{split}
 \langle \psi_\ua(\rv)^\dagger \psi_\ua(\zerov)\rangle 
 &\approx n \exp \left\{ -\frac12 \langle \left[ \theta_\ua(\rv)-\theta_\ua(\zerov) \right]^2\rangle \right\}\\
 &= n \exp\left\{ \frac{1}{nF_1} \left[ U_d(\rv;\alpha)-U_d(\zerov;\alpha) \right] \right\}.
\end{split}
\end{equation}
For $d=1,2,3$, we specifically have
\begin{subequations}\label{eq:corr_psi_d123}
\begin{align}
 d=1:&~\langle \psi_\ua(x)^\dagger \psi_\ua(0)\rangle \approx n\left( \frac{D(x|L)}{\alpha} \right)^{-\frac{1}{\pi nF_1}}~~(|x|\gg\alpha); \label{eq:corr_psi_d1}\\
 d=2:&~\langle \psi_\ua(\rv)^\dagger \psi_\ua(\zerov)\rangle \approx n\exp \left[\frac{1}{2\pi nF_1} \left( \frac{1}{\sqrt{r^2+\alpha^2}}-\frac{1}{\alpha} \right) \right];\\
 d=3:&~\langle \psi_\ua(\rv)^\dagger \psi_\ua(\zerov)\rangle \approx n\exp \left[\frac{1}{2\pi^2 nF_1} \left( \frac{1}{r^2+\alpha^2}-\frac{1}{\alpha^2} \right) \right].
\end{align}
\end{subequations}
Here, $D(x|L)$ is the chord distance defined in Eq.\ \eqref{eq:chord} \footnote{
The power-law decay in terms of the chord distance as in Eq.\ \eqref{eq:corr_psi_d1}
is well-known in the behavior 
of correlation functions of primary fields in conformal field theory. See, e.g., Appendix C of Ref.\ \cite{Cazalilla04}.
}.
We note that use of the convergence factor in the present approach leaves the ambiguity of $\alpha$ in the final results,  
and that a more precise calculation of this correlation function requires a careful analysis of 
the phase correlation $\langle \theta_{-\kv,\ua}\theta_{ \kv,\ua} \rangle$ up to large $k$ \cite{Khawaja02,Mora03}. 
Yet, the present simple approach can still provide a qualitative picture. 

The results in Eq.\ \eqref{eq:corr_psi_d123} show a quasi-long-range order for $d=1$ and a long-range order (LRO) for $d\ge 2$. 
The long-range interaction in terms of the superfluid velocity (i.e., the $F_1$ term) in the entanglement Hamiltonian \eqref{eq:He_Omega_Real}
plays a key role in the emergence of this (quasi-)LRO in a canonical ensemble 
$\langle\cdot\rangle=\Tr \left(\cdot ~\erm^{-H_\erm}\right) / \Tr~ \erm^{-H_\erm}$, especially, for $d=1$ and $2$. 
If $H_\erm$ did not have such a term but had a form similar to a scalar Bose gas (i.e., the spin-$\ua$ part of the original Hamiltonian), 
the one-particle density matrix would show an exponential decay for $d=1$ \cite{Cazalilla04} and a quasi-LRO for $d=2$ \cite{Khawaja02} at nonzero fictitious temperatures. 
Equation \eqref{eq:corr_psi_d123} also indicates that an increase in the coefficient $F_1$ 
leads to a slower spatial decay of the one-particle density matrix, which can be understood as a consequence of the suppression of the phase fluctuation discussed above. 
For $d=1$, in particular, $F_1$ appears in the decay exponent $1/(\pi nF_1)$. 

\subsection{Case of $\Omega=0$}

Using the gapped ES $\xi_\kv$ in Eq.\ \eqref{eq:xi_k_Omega=0} and the coefficients $F_\kv$ and $G_\kv$ in Eq.\ \eqref{eq:Fk_Gk_withoutRabi}, 
we can calculate the phase and density fluctuations \eqref{eq:Correlator_FkGk} in the long-wavelength limit as 
\begin{subequations}\label{eq:corr_theta_n_Omega=0_FG}
\begin{align}
 \langle \theta_{-\kv,\ua}\theta_{ \kv,\ua} \rangle &
  \approx \frac{1}{nk} \left( \frac{G_{-1}}{F_1} \right)^{1/2} \left[f_\Brm(\sqrt{F_1G_{-1}})+\frac12\right] \notag\\
  &= \frac{\gt_+^{1/2}+\gt_-^{1/2}}{8nk} ,\label{eq:corr_theta_Omega=0}\\
 \langle n_{-\kv,\ua} n_{\kv,\ua} \rangle &
  \approx nk \left( \frac{F_1}{G_{-1}} \right)^{1/2} \left[f_\Brm(\sqrt{F_1G_{-1}})+\frac12\right]\notag\\
  &= \frac{n}{2} \left( \gt_+^{-1/2}+\gt_-^{-1/2} \right) k ~~(\kv\ne\zerov). \label{eq:corr_n_Omega=0}
\end{align}
\end{subequations}
We notice that these correlations are related to the coefficients $F_1$ and $G_{-1}$ in the entanglement Hamiltonian
in a manner more complicated than in Eq.\ \eqref{eq:corr_theta_n_Omega>0_FG}; 
yet, we again find that an increase in $F_1$ ($G_{-1}$) leads to a suppression of the phase (density) fluctuation.  
Furthermore, different dependences of $\langle n_{-\kv,\ua} n_{\kv,\ua} \rangle$ on $k$ in Eqs.\ \eqref{eq:corr_theta_n_Omega>0_G} and \eqref{eq:corr_n_Omega=0} 
indicate that in the long-wavelength limit, the density fluctuation is qualitatively more suppressed for $\Omega=0$ than for $\Omega>0$;  
for $d=1$, this suppression leads to a quasi-LRO in the string correlation as we explain later. 

Equation \eqref{eq:corr_theta_Omega=0} indicates
\begin{equation}\label{eq:corr_vs_Omega=0}
 \bra{0^\osc} (-\irm\kv \theta_{-\kv,\ua})\cdot (\irm\kv\theta_{ \kv,\ua}) \ket{0^\osc} \approx  \frac{\gt_+^{1/2}+\gt_-^{1/2}}{8n} k .
\end{equation}
The fact that both Eqs.\ \eqref{eq:corr_n_Omega=0} and \eqref{eq:corr_vs_Omega=0} show linear behavior for small $k$ 
explains why the same type of long-range interactions appear 
in the superfluid velocity $\vv_{\srm,\ua}=-\frac{\hbar}{M}\nabla\theta_\ua$ and the density  $n_\ua$
in the entanglement Hamiltonian \eqref{eq:He_Omega_Real_Omega=0}. 
For $d=1$, a similar behavior in terms of $\partial_x\theta_\ua$ and $n_\ua-n$ 
can be understood as a consequence of the duality between the two fields $\phi_\pm/\sqrt{K_\pm}$ and $\sqrt{K_\pm} \vartheta_\pm$ 
in the TLL Hamiltonian \eqref{eq:H_2TLL} with $m_\pm^2=0$. 

Using Eq.\ \eqref{eq:corr_theta_Omega=0}, the one-particle density matrix can be calculated in the same manner as in Sec.\ \ref{sec:corr_Omega>0}. 
The resulting expressions are given by Eqs.\ \eqref{eq:corr_psi_Ud} and \eqref{eq:corr_psi_d123} 
with the replacement of $1/F_1$ by $\left(\gt_+^{1/2}+\gt_-^{1/2}\right)/8$. 
We thus again have a quasi-LRO for $d=1$ and a LRO for $d\ge 2$, 
for which the long-range interaction in terms of $\nabla\theta_\ua$ in $H_\erm$ plays a crucial role. 
We note that for $d=1$, the decay exponent is given by $\left(\gt_+^{1/2}+\gt_-^{1/2}\right)/(8\pi n)=(K_+^{-1}+K_-^{-1})/4$, 
where $K_\pm$ are introduced in Eq.\ \eqref{eq:H_2TLL_corres}. 

It is then natural to ask what role the long-range interaction in terms of $n_\ua-n$ in $H_\erm$ plays in the correlation properties. 
To answer this question, we focus on the 1D case, and introduce the field $\phi_\ua(x)$ such that $\partial_x \phi_\ua=\sqrt{\pi}(n_\alpha-n)$. 
In analogy with the phase part of Eq.\ \eqref{eq:corr_psi_theta_n}, we consider 
\begin{equation}\label{eq:corr_phi_string}
\begin{split}
 &\left\langle \erm^{\irm 2\sqrt{\pi} \lambda \left( \phi_\ua(x)-\phi_\ua(0) \right)} \right\rangle\\
 &= \left\langle \exp \left[ \irm 2\pi \lambda \int_0^x \drm x' \left(n_\alpha(x')-n\right)\right]\right\rangle ,
\end{split}
\end{equation}
where $\lambda$ is a real constant. 
This can be viewed as a string order parameter as it involves the integration of the density over the interval $[0,x]$. 
It can also be viewed as a characteristic function of the atom-number statistics in the interval $[0,x]$ \cite{Song10}. 
For $\lambda=1/2$, this quantity has been used to characterize the Mott insulator phase in optical lattices \cite{Berg08} 
and measured experimentally \cite{Endres11}. 
To calculate Eq.\ \eqref{eq:corr_phi_string}, we first use Eq.\ \eqref{eq:corr_n_Omega=0} 
to obtain the fluctuation of the Fourier component $\phi_{k,\ua}$ of $\phi_\ua(x)$ as
\begin{equation}\label{eq:corr_phi_Omega=0}
 \langle \phi_{-k,\ua} \phi_{k,\ua} \rangle 
 \approx \frac{\pi n}{2 k} \left( \gt_+^{-1/2}+\gt_-^{-1/2} \right)=\frac{K_++K_-}{4k}  ~~(k\ne 0). 
\end{equation}
Following a similar line of calculations as in Sec.\ \ref{sec:corr_Omega>0}, we obtain
\begin{equation}\label{eq:corr_phi_string_qLRO}
\begin{split}
 &\left\langle \erm^{\irm 2\sqrt{\pi} \lambda \left( \phi_\ua(x)-\phi_\ua(0) \right)} \right\rangle\\
 &=\exp \left\{ -2\pi\lambda^2 \left\langle \left[ \phi_\ua(x)-\phi_\ua(0) \right]^2  \right\rangle \right\}\\
 &\approx \left( \frac{D(x|L)}{\alpha} \right)^{-\lambda^2(K_++K_-)}~~(|x|\gg\alpha).
\end{split}
\end{equation}
We thus obtain a quasi-LRO in the string correlation. 
From the viewpoint of the entanglement Hamiltonian \eqref{eq:He_Omega_Real_Omega=0}, 
this is a consequence of the suppression of the density fluctuation due to the long-range interaction in terms of the density. 


\section{Summary and outlook}\label{sec:summary}

We have studied the intercomponent ES in  binary BECs in $d$ spatial dimensions. 
By means of effective field theory, we have shown that the ES exhibits an anomalous square-root dispersion relation \eqref{eq:xi_k_Omega>0} 
for a finite Rabi coupling $\Omega>0$ and a gapped dispersion relation \eqref{eq:xi_k_Omega=0} in its absence ($\Omega=0$). 
We have related these intriguing spectra to the emergence of long-range interactions in terms of the superfluid velocity and the particle density 
in the entanglement Hamiltonian [see Eqs.\ \eqref{eq:He_Omega_Real} and \eqref{eq:He_Omega_Real_Omega=0}]. 
We have discussed how these unusual interactions manifest themselves in the properties of intracomponent correlation functions. 
Using the obtained ES, we have also calculated the intercomponent EE. 
The result for $\Omega>0$ in Eq.\ \eqref{eq:Se_Omega>0} 
shows a volume-law scaling followed by subleading logarithmic terms. 
The coefficient of the logarithmic contribution is $(d-1)/2$ in total, 
where $d/2$ originates from the restoration of the global U$(1)$ symmetry for a finite volume 
and $-1/2$ from the Nambu-Goldstone mode. 
The result for $\Omega=0$ in Eq.\ \eqref{eq:Se_Omega=0} shows a volume-law scaling accompanied by a negative universal constant $-s_0$. 
Here, the constant $s_0$ is given by a universal function \eqref{eq:s0_Omega=0} of the ratio of 
the effective coupling constants $g_\pm$ in the symmetric and antisymmetric channels. 

It is of interest to extend the present work to other types of multicomponent systems such as spinor BECs \cite{Kawaguchi12} and spin-orbit coupled gases \cite{Lin11,Zhai12_review,Po14,Yoshino19}. 
In particular, Po {\it et al.}\ \cite{Po14} have shown the emergence of a non-TLL quantum liquid with a quadratic energy dispersion relation 
in 1D spin-orbit-coupled Bose gases. It is worth examining how the intercomponent ES behaves in this unusual case. 
We can further expect rich behavior in two coupled systems with higher continuous symmetry beyond U$(1)$, 
where the intercomponent coupling can be expressed as a generalized Josephson coupling \cite{Beekman20}. 

\bigskip
We are grateful to Taiki Haga for stimulating discussions. 
This work was supported by KAKENHI Grant No. JP18H01145 and No. JP18K03446, 
a Grant-in-Aid for Scientific Research on Innovative Areas ``Topological Materials Science'' (KAKENHI Grant No. JP15H05855) 
from the Japan Society for the Promotion of Science (JSPS), and Keio Gijuku Academic Development Funds.
TY was supported by JSPS through the Program for Leading Graduate School (ALPS). 

\appendix
\section{Long-range interaction potential $U_d(\rv-\rv')$}\label{App:derivation_F_d}

For $d$ spatial dimensions, we have introduced the long-range interaction potential $U_d(\rv-\rv')$ in Eq.\ \eqref{eq:F_d_Bare}, 
which emerges in the entanglement Hamiltonian. 
Here we derive its expressions \eqref{eq:U123} for $d=1,2,3$. 
In the following, we set $\rv'=\zerov$ and calculate
\begin{equation}\label{eq:F_d_r_definition}
 U_d(\rv;\alpha)=\sum_{\kv\neq\zerov} \frac{1}{V|\kv|} \erm^{-\alpha|\kv|+\irm\kv\cdot\rv}.
\end{equation}
For $d>1$, we can take the infinite-volume limit $V\to\infty$ to rewrite this in the integral form as
\begin{equation}\label{eq:Ud_integral}
 U_d(\rv;\alpha)=\int \frac{\mathrm{d}^d\kv}{(2\pi)^d} \frac{1}{|\kv|} \erm^{-\alpha|\kv|+\irm\kv\cdot\rv}.
\end{equation}
After calculating Eq.\ \eqref{eq:F_d_r_definition} or Eq.\ \eqref{eq:Ud_integral}, 
we take the limit $\alpha\to 0^+$ to obtain $U_d(\rv)$. 

For $d=1$, the sum in Eq.\ \eqref{eq:F_d_r_definition} is taken as
\begin{equation}
\begin{split}
 U_1(x;\alpha)
 &=\sum_{k\neq0} \frac{1}{L|k|} \erm^{-\alpha|k|+\irm kx} \\
 &= \frac{1}{2\pi} \sum_{n=1}^\infty \frac1n \left[ \erm^{\frac{2\pi}{L} (-\alpha+\irm x) n} +\mathrm{c.c.} \right]\\
 &=-\frac{1}{2\pi} \ln  \left|1-\erm^{\frac{2\pi}{L} (-\alpha+\irm x) }\right|^2,
\end{split}
\end{equation}
where we use $\sum_{n=1}^\infty z^n/n =\mathrm{Li}_1(z)= -\ln (1-z)~(|z|<1)$. 
Taking $\alpha\to 0^+$, we obtain
\begin{equation}
\begin{split}
 U_1(x) = -\frac{1}{\pi} \ln \left|\erm^{\irm \frac{2\pi}{L}x}-1\right| 
 = -\frac{1}{\pi}\ln\frac{2\pi D(x-x'|L)}{L},
\end{split}
\end{equation}
where $D(x-x'|L)$ is the chord distance \eqref{eq:chord}. 

For $d=2$, we can introduce the polar coordinate $(k,\theta)$ to rewrite the integral \eqref{eq:Ud_integral} as
\begin{equation}\label{eq:U2_integral}
\begin{split}
 U_2(\rv;\alpha)
 &=\frac{1}{(2\pi)^2} \int_0^{2\pi}\mathrm{d}\theta \int_0^\infty \mathrm{d}k\ \erm^{-\alpha k+\irm kr\cos\theta}  \\
 &=\frac{1}{(2\pi)^2} \int_0^{2\pi}\mathrm{d}\theta \frac{1}{ \alpha-ir\cos\theta } .
\end{split}
\end{equation}
Defining $z=\erm^{\irm\theta}$, the last integral can be rewritten as a contour integral along the unit circle: 
\begin{equation}
\begin{split}
 U_2(\rv;\alpha)  
 &= \frac{1}{2\pi^2}\oint \mathrm{d}z \frac{1}{r(z^2+1)+2\mathrm{i}\alpha z}\\
 &= \frac{1}{2\pi^2 r}\oint \mathrm{d}z \frac{1}{(z-z_+)(z-z_-)}, 
\end{split}
\end{equation}
where $z_\pm=\irm\left(-\alpha\pm\sqrt{r^2+\alpha^2} \right)/r$ are the locations of poles. 
Since $|z_+|<1<|z_-|$, the integral picks up only the residue at $z=z_+$, leading to 
\begin{equation}
 U_2(\rv;\alpha)  =\frac{\irm}{\pi r(z_+-z_-)} = \frac1{2\pi\sqrt{r^2+\alpha^2}} \to \frac{1}{2\pi r} ~(\alpha\to 0^+).  
\end{equation}

For $d=3$, we can introduce the polar coordinate $(k,\theta,\phi)$ to perform the integral \eqref{eq:Ud_integral} as
\begin{equation}\label{eq:U3_integral}
\begin{split}
 U_3(\rv;\alpha)
 =&\frac{1}{(2\pi)^3} \int_0^{2\pi}\!\! \mathrm{d}\phi \int_{-1}^{1}\!\!\mathrm{d}\cos\theta \int_0^\infty\!\! \mathrm{d}k \ 
    k \erm^{-\alpha k+\irm kr\cos\theta} \\
 =&\frac{1}{(2\pi)^2 \irm r} \int_0^\infty \!\!\mathrm{d}k \left(\erm^{-\alpha k+\irm kr}-\erm^{-\alpha k-\irm kr}\right) \\
 =&\frac{1}{2\pi^2(r^2+\alpha^2)} 
 \to \frac{1}{2\pi^2r^2}~(\alpha\to 0^+). 
\end{split}
\end{equation}

\section{Oscillator-mode contribution to the entanglement entropy}\label{App:EE}

Here we calculate the oscillator-mode contribution $S_\mathrm{e}^\osc$ to the EE 
for the cases in which the ES shows gapless and gapped dispersion relations $\xi_\kv$ as in Eqs.\ \eqref{eq:xi_k_Omega>0} and \eqref{eq:xi_k_Omega=0}.
To this end, it is useful to consider the canonical ensemble given by $H_\mathrm{e}^\osc$ at the fictitious temperature $T$. 
In this ensemble, the number operator $\eta_\kv^\dagger \eta_\kv $ appearing in Eq.~\eqref{eq:DiagonalHam} obeys 
the Bose distribution $\langle \eta_\kv^\dagger \eta_\kv \rangle=\left( e^{\xi_\kv/T}-1 \right)^{-1}=f_\Brm \left(\xi_\kv/T\right)$, from which the internal energy can be calculated.
We can then apply the thermodynamic relation to calculate the thermal entropy, and obtain the EE $S_\mathrm{e}^\osc$ by setting $T=1$. 

\subsection{Case of the gapped ES}\label{App:EE_gapped}

We first consider the case in which the ES shows the gapped dispersion relation $\xi_\kv = \xi_0+c k^\gamma$ with $\xi_0,c,\gamma>0$. 
The internal energy $E_\mathrm{e}^\osc(T)$ (measured relative to the zero-point energy) at the fictitious temperature $T$ is calculated as 
\begin{equation}\label{eq:Ee_osc_gapped}
\begin{split}
  E_\mathrm{e}^\osc(T)
 &= \sum_{\kv\ne \zerov} \xi_\kv f_\Brm \left(\xi_\kv/T\right) \\
 &= \sum_\kv \xi_\kv f_\Brm\left(\xi_\kv/T\right)- \xi_0 f_\Brm\left(\xi_0/T\right) \\
 &= e(T) V-\xi_0 f_\Brm\left(\xi_0/T\right),
\end{split}
\end{equation}
where 
\begin{equation}\label{eq:eT_gapped}
\begin{split}
 e(T) &=\int \frac{\drm^d\kv}{(2\pi)^d} \frac{\xi_\kv}{\erm^{\xi_\kv/T}-1}\\
 &= \frac{S_d}{(2\pi)^d} \int_0^\infty k^{d-1}\drm k \frac{\xi_0+c k^\gamma}{\erm^{(\xi_0+c k^\gamma)/T}-1}\\
 &= \frac{S_d }{(2\pi)^d\gamma c^{d/\gamma}} 
 \bigg[  \Gamma\left(\frac{d}{\gamma}\right) \mathrm{Li}_{d/\gamma} \left(\erm^{-\frac{\xi_0}{T}}\right) \xi_0 T^{d/\gamma}\\
 &~~~~~+\Gamma\left(\frac{d}{\gamma}+1\right)\mathrm{Li}_{d/\gamma+1} \left(\erm^{-\frac{\xi_0}{T}}\right) T^{d/\gamma+1}\bigg].
\end{split}
\end{equation}
Here, $S_d:=2\pi^{d/2}/\Gamma(d/2)$ is the surface area of the unit $d$-sphere. 
The EE is then calculated as
\begin{equation}\label{eq:Se_osc_gapped}
 S_\erm^\osc=\int_0^1 \frac{\drm T}{T} \left( \frac{\partial E_\erm^\osc}{\partial T} \right)_V =s_1 V-s_0,
\end{equation}
where
\begin{equation}\label{eq:Se_osc_gapped_coeff}
 s_1=\int_0^1 \frac{\drm T}{T} \frac{\drm e}{\drm T},~~s_0=\frac{\xi_0}{\erm^{\xi_0}-1}-\ln \left( 1-\erm^{-\xi_0}\right) .
\end{equation}

\subsection{Case of the gapless ES}\label{App:EE_gapless}

We next consider the case in which the ES shows the gapless dispersion relation $\xi_\kv = c k^\gamma$ with $c,\gamma>0$. 
A calculation similar to Eq.~\eqref{eq:Ee_osc_gapped} yields 
\begin{equation}\label{eq:Ee_osc_gapless}
 E_\erm^\osc (T)=\frac{S_d \Gamma(d/\gamma+1)\zeta(d/\gamma+1)}{(2\pi)^d \gamma}  \frac{VT^{d/\gamma+1}}{c^{d/\gamma}} -T.
\end{equation}
The second term $-T$ gives a divergent contribution to $S_\erm^\osc$ when performing an integration over $T\in [0,1]$ as in Eq.\ \eqref{eq:Se_osc_gapped}. 
Since Eq.\ \eqref{eq:Ee_osc_gapless} is based on the approximation of the $\kv$ sum by integration, 
it may be invalidated at low $T$ where the discrete nature of $\kv$ becomes important; 
the above divergence can be interpreted as a consequence of using such an invalid expression at low $T$. 
Yet, we can still use Eq.\ \eqref{eq:Ee_osc_gapless} in an integral over $T$ above a certain temperature, 
and find a contribution $-\ln T$ to the thermal entropy $S_\erm^\osc(T)$. 
This contribution should appear in the form $-\ln (L^\gamma T/c)$ as the entire energy spectrum is proportional to $c/L^\gamma$. 
By including the extensive contribution from the first term of Eq.\ \eqref{eq:Ee_osc_gapless} and setting $T=1$, we obtain 
\begin{equation}\label{eq:Se_osc_gapless}
 S_\erm^\osc= \frac{\sigma_d L^d}{c^{d/\gamma}}-\ln \frac{L^\gamma}{c}+O(1),
\end{equation}
where
\begin{equation}\label{eq:Se_osc_gapless_coeff}
 \sigma_d:=\frac{S_d \Gamma(d/\gamma+2)\zeta(d/\gamma+1)}{(2\pi)^d d }.
\end{equation}

The logarithmic term in Eq.\ \eqref{eq:Se_osc_gapless} can also be obtained in the following way. 
We first express $S_\erm^\osc$ as a discrete sum
\begin{equation}\label{eq:Se_osc_gapless_1}
 S_\erm^\osc = \sum_{\kv\ne \zerov} \left[ \frac{\xi_\kv}{\erm^{\xi_\kv}-1}-\ln\left( 1-\erm^{-\xi_\kv}\right) \right],
\end{equation}
where the summand is the entropy of a harmonic oscillator with the energy-level spacing $\xi_\kv$ at $T=1$. 
We introduce an ultraviolet cutoff for the wave vector $\kv=(k_1,\dots,k_d)$ in such a manner that
each element $k_i$ runs over $k_i=2\pi n_i/L$ with $n_i\in\{-\Lambda,-\Lambda+1,\dots, \Lambda-1\}$, where $\Lambda$ is proportional to $L$.  
Then Eq.~\eqref{eq:Se_osc_gapless_1} becomes a finite sum. 
We can rewrite it as
\begin{equation}\label{eq:Se_osc_gapless_2}
 S_\erm^\osc = \sum_{\kv\ne \zerov} \left[ \frac{\xi_\kv}{\erm^{\xi_\kv}-1}-\ln\frac{1-\erm^{-\xi_\kv}}{\xi_\kv}\right] - \sum_{\kv\ne\zerov} \ln \xi_\kv ,
\end{equation}
and treat the first and second sums separately. 
In the first sum, the summand is convergent in the limit $\kv\to\zerov$, 
and we can apply the Euler-Maclaurin formula (at the order of the trapezoid formula) $d$ times to rewrite this part as
\begin{equation}
 L^d \int_{-\lambda}^\lambda \frac{\drm^d \kv}{(2\pi)^d} \left[ \frac{\xi_\kv}{\erm^{\xi_\kv}-1}-\ln\frac{1-\erm^{-\xi_\kv}}{\xi_\kv}\right]-1,
\end{equation}
where $\lambda:=2\pi\Lambda/L$. 
The second sum in Eq.\ \eqref{eq:Se_osc_gapless_2} requires a careful treatment as the summand diverges for $\kv\to\zerov$. 
We can rewrite this part as
\begin{equation}
 -\sum_{\kv\ne\zerov}\ln \xi_\kv = - \left[ \left(\frac{\lambda L}{\pi}\right)^d-1\right]\ln c -\frac{\gamma}{2} \sum_{\kv\ne\zerov} \ln \kv^2,
\end{equation}
and focus on the calculation of $\sum_{\kv\ne\zerov}\ln\kv^2$. 
Introducing $\kv_\perp=(k_2,\dots,k_d)$ and $\delta=2\pi/L$ and applying the Euler-Maclaurin formula, we have (see Ref.\ \cite{Misguich17} for a related calculation)
\begin{equation}\label{eq:lnk2}
\begin{split}
 &\sum_{\kv\ne\zerov} \ln (\kv^2) \\
 &= \sum_{\kv_\perp \ne\zerov}\sum_{k_1} \ln(k_1^2+\kv_\perp^2) + \sum_{k_1\ne 0}\ln k_1^2\\
 &= \frac{L}{2\pi} \sum_{\kv_\perp \ne\zerov} \int_{-\lambda}^\lambda \drm k_1 \ln (k_1^2+\kv_\perp^2) + \frac{L}{\pi} \int_{\delta}^\lambda \drm k_1 \ln k_1^2 + \ln \delta^2\\
 &=\frac{L}{2\pi} \sum_{\kv_\perp \ne\zerov} \left[2\lambda \ln(\lambda^2+\kv_\perp^2)+4|\kv_\perp| \arctan \frac{\lambda}{|\kv_\perp|}-4\lambda\right] \\
 &~~~+ \frac{2L}{\pi} \left( \lambda \ln\lambda-\lambda-\delta\ln\delta+\delta \right) +\ln\delta^2\\
 &=\frac{L^d}{(2\pi)^d} \int_{[-\lambda,\lambda]^{d-1}} \drm^{d-1}\kv_\perp \bigg[2\lambda \ln(\lambda^2+\kv_\perp^2)\\
 &~~~~~~~~~~~~~~+4|\kv_\perp| \arctan \frac{\lambda}{|\kv_\perp|}-4\lambda\bigg]  -2\ln\delta +4.
\end{split}
\end{equation}
Combining these results, we obtain 
\begin{equation}\label{eq:Se_osc_gapless_EM}
 S_\erm^\osc =  \frac{\sigma L^d}{c^{d/\gamma}} -\ln \frac{L^\gamma}{(2\pi)^\gamma c}-2\gamma-1,
\end{equation}
where the coefficient $\sigma$ of the extensive part depends on $\lambda$ or, more generally, on how we introduce the ultraviolet cutoff. 
In the limit $\lambda\to\infty$, $\sigma$ is expected to converge to $\sigma_d$ in Eq.\ \eqref{eq:Se_osc_gapless_coeff}. 

Focusing on the case of $\gamma=1/2$, which is relevant to the case of Sec.\ \ref{sec:ES_Omega>0}, 
we have performed a numerical check of Eq.\ \eqref{eq:Se_osc_gapless_EM}. 
We set $c=1$, vary $L$ over $L\in\{100,200,\dots,1000\}$, and calculate the finite sum in Eq.\ \eqref{eq:Se_osc_gapless_1}. 
We fit the obtained data to the form $S_\erm^\osc=\sigma L^d-\beta \ln \sqrt{L/(2\pi)}-s_0$. 
For $d=1$, we obtain the following results for $\lambda/(2\pi)=\Lambda/L=10$ and $100$: 
\begin{equation}
\begin{split}
\frac{\lambda}{2\pi}=10:&~\sigma=2.27516,~\beta=0.99938,~s_0=1.92018;\\
\frac{\lambda}{2\pi}=100:&~\sigma=2.29576,~\beta=0.99938,~s_0=1.92018.
\end{split}
\end{equation}
For $d=2$, we obtain the following results for $\lambda/(2\pi)=5$ and $10$: 
\begin{equation}
\begin{split}
\frac{\lambda}{2\pi}=5:&~\sigma=7.32834,~\beta=0.99935,~s_0=1.68177;\\
\frac{\lambda}{2\pi}=10:&~\sigma=9.21249,~\beta=0.99935,~s_0=1.66113.
\end{split}
\end{equation}
In both cases, the coefficient $\beta$ of the logarithmic term agrees excellently with the expected value $\beta=1$. 
As for the leading extensive contribution, the coefficient $\sigma$ monotonically increases with an increase in $\lambda$. 
For $d=1$ and $2$, this coefficient is expected to converge, in the $\lambda\to\infty$ limit, 
to  $\sigma_1=2.29576$ and $\sigma_2=9.90193$, respectively, which are given by Eq.\ \eqref{eq:Se_osc_gapless_coeff}; 
for $d=1$, we find a good convergence to this value already for $\lambda/(2\pi)=100$. 
The constant term $s_0$ obtained numerically deviates from the expected value $s_0=2$ in Eq.\ \eqref{eq:Se_osc_gapless_EM}; 
even if we change the range of $L$ to larger values, we find that this constant stays almost constant (not shown). 
We therefore conclude that Eq.\ \eqref{eq:Se_osc_gapless_EM} obtained from the Euler-Maclaurin formula at the order of the trapezoid formula 
acquires a constant correction beyond this order. 

\section{Effect of a symmetry-breaking field}\label{app:symbr}

\newcommand{\Deltat}{\tilde{\Delta}}

Here we discuss the effect of a small symmetry-breaking field. 
Specifically, we add the term $\sqrt{n} h \sum_\alpha \left( \psi_\alpha + \psi_\alpha^\dagger -2\sqrt{n} \right)$ 
with $h>0$ to the Lagrangian density \eqref{eq:Lag_theta_n_pre}. 
In this case, the total number of particles, $N$, is no longer conserved, 
and we have to introduce the chemical potential $\mu=g_+n-\frac{\hbar\Omega}{2}-h$ to have the desired density $n$ for each component. 
The Hamiltonian \eqref{eq:H_theta_n_k} with the symmetry-breaking and chemical potential terms is expressed as 
\begin{equation}\label{eq:H_theta_dn}
\begin{split}
 H-\mu N
 =&\sum_{\kv} \sum_{\nu=\pm}
   \bigg[\frac{n}{2} (\epsilon_\kv+h+\hbar\Omega\delta_{\nu,-})\theta_{-\kv,\nu}\theta_{\kv,\nu}\\
 &~~~+\frac1{2n} (\epsilon_\kv+h+2g_{\nu}n) \delta n_{-\kv,\nu} \delta n_{\kv,\nu} \bigg] ,
\end{split}
\end{equation}
where we define $\delta n_{\kv,\nu}:=n_{\kv,\nu}-\sqrt{V}n\delta_{\kv,\zerov}\delta_{\nu,+}$ and ignore unimportant constant terms. 
Because of the $h$ term, the coefficient of $\theta_{-\kv,\nu}\theta_{\kv,\nu}$ in the Hamiltonian is positive for all $\kv$, 
and we can treat the zero mode in the same way as the oscillator mode. 
We introduce annihilation and creation operators $\gamma_{\kv,\nu}$ and $\gamma_{\kv,\nu}^\dagger$ for all $\kv$ 
in a manner similar to Eq.\ \eqref{eq:gamma_k} but we replace $n_{\kv,\nu}$ by $\delta n_{\kv,\nu}$. 
Here, the factor $\zeta_{\kv,\nu}$ in Eqs.\ \eqref{eq:zeta_pm} and \eqref{eq:zeta_pm_simp} is changed to 
\begin{equation}\label{eq:zeta_pm_Delta}
 \zeta_{\kv,\nu}
 =\left(\frac{\epsilon_\kv+h+\hbar\Omega\delta_{\nu,-}}{\epsilon_\kv+h+2g_{\nu}n}\right)^{1/4}
 =\left( \frac{\Omegat\delta_{\nu,-}+k^2+\tilde{h}}{\gt_\nu+k^2+\tilde{h}} \right)^{1/4},
\end{equation}
where $\tilde{h}:=2Mh/\hbar^2$. 
The Hamiltonian \eqref{eq:H_theta_dn} is then rewritten as the sum of harmonic oscillators similar to Eq.\ \eqref{eq:decouple Hamiltonian}  
but the sum now includes $\kv=\zerov$ as well. 
The excitation spectrum \eqref{eq:E_k} is changed to 
\begin{equation}
 E_\nu(\kv)
 =\sqrt{\left(\epsilon_\kv+h+\hbar\Omega\delta_{\nu,-}\right)
   \left(\epsilon_\kv+h+2g_{\nu}n\right)} .
\end{equation}
This shows that the symmetric channel is also gapped with the excitation energy $E_+(\zerov)=\sqrt{2g_+n h}$. 
In the following, we assume $\tilde{h}\ll \left( 2\pi/L \right)^2$ so that the effect of $h$ is negligible for $\kv\ne\zerov$. 
We also assume that $E_+(\zerov)=\sqrt{2g_+nh}$ is much larger than the energy spacing $g_+/(4V)$ for $h=0$ in Eq.\ \eqref{eq:Hzero_spec} 
so that the effect of $h$ is significant in the zero mode. 

Let us now discuss the entanglement properties. 
The entanglement Hamiltonian has the form of Eq.\ \eqref{eq:DiagonalHam} with the modification of including $\kv=\zerov$. 
The single-particle ES $\xi_\kv$ is still given by Eq.\ \eqref{eq:xi_general} but the factors $\zeta_{\kv,\pm}$ are replaced by Eq.\ \eqref{eq:zeta_pm_Delta}. 
A simple method for incorporating the effect of $h$ into the results in Sec.\ \ref{sec:ES} is 
to perform the replacement $k\to \sqrt{k^2+\tilde{h}}$ 
and then to expand the resulting expressions in terms of $\tilde{h}$ when necessary. 

We first consider the case of $\Omega>0$. 
For $\kv\ne\zerov$, the single-particle ES is still given by $\xi_\kv\approx c_{1/2}k^{1/2}$ as in Eq.\ \eqref{eq:xi_k_Omega>0}. 
Thus, the contribution of the $\kv\ne\zerov$ modes to the EE in Eq.\ \eqref{eq:Se_osc_Omega>0} remains unchanged. 
For $\kv=\zerov$, we have $\xi_\zerov \approx c_{1/2} \tilde{h}^{1/4}$. 
Therefore, the zero-mode contribution to the EE is no longer given by Eq.\ \eqref{eq:Se_zero_Omega>0} but is replaced by  
\begin{equation}\label{eq:EE_zero_ht}
 S_\mathrm{e}^\zero = \frac{\xi_\zerov}{\erm^{\xi_\zerov}-1}-\ln \left( 1-\erm^{-\xi_\zerov}\right) 
 \approx \ln \frac{\erm}{c_{1/2} \tilde{h}^{1/4}} .
\end{equation}
We thus find that the EE depends logarithmically on $h$. 
A similar logarithmic dependence on a symmetry-breaking field has also been found in the subregion EE \cite{Laflorencie16,Metlitski11,Song11,Laflorencie15,Luitz15ssb}. 
In the present argument, we have assumed $E_+(\zerov)=\sqrt{2g_+nh}\gg g_+/(4V)$, i.e., $\tilde{h}\gg \gt_+/(64n^2V^2)$. 
If we tune $\tilde{h}$ to this cutoff in Eq.\ \eqref{eq:EE_zero_ht}, we obtain 
\begin{equation}\label{eq:EE_zero_ht_V}
 S_\mathrm{e}^\zero \approx \frac12 \ln \frac{\erm^2 nV}{G_0}, 
\end{equation}
which shows the same logarithmic dependence on $V$ as Eq.\ \eqref{eq:Se_zero_Omega>0}. 
This is consistent with the argument in Refs.\ \cite{Song11,Laflorencie15,Luitz15ssb} 
that calculations with a properly chosen symmetry-breaking field can mimic 
the behavior of the symmetry-restored ground state. 

We next consider the case of $\Omega=0$. 
The single-particle ES in Eq.\ \eqref{eq:xi_k_Omega=0} can now be used for $\kv=\zerov$ as well. 
Therefore, the universal constant term $-s_0$ in Eq.\ \eqref{eq:Se_Omega=0} disappears. 
We thus realize that the restoration of the U$(1)\times$U$(1)$ symmetry for a finite volume is crucial for the emergence of the universal constant. 




\bibliography{reference}

\begin{thebibliography}{104}%
\makeatletter
\providecommand \@ifxundefined [1]{%
 \@ifx{#1\undefined}
}%
\providecommand \@ifnum [1]{%
 \ifnum #1\expandafter \@firstoftwo
 \else \expandafter \@secondoftwo
 \fi
}%
\providecommand \@ifx [1]{%
 \ifx #1\expandafter \@firstoftwo
 \else \expandafter \@secondoftwo
 \fi
}%
\providecommand \natexlab [1]{#1}%
\providecommand \enquote  [1]{``#1''}%
\providecommand \bibnamefont  [1]{#1}%
\providecommand \bibfnamefont [1]{#1}%
\providecommand \citenamefont [1]{#1}%
\providecommand \href@noop [0]{\@secondoftwo}%
\providecommand \href [0]{\begingroup \@sanitize@url \@href}%
\providecommand \@href[1]{\@@startlink{#1}\@@href}%
\providecommand \@@href[1]{\endgroup#1\@@endlink}%
\providecommand \@sanitize@url [0]{\catcode `\\12\catcode `\$12\catcode
  `\&12\catcode `\#12\catcode `\^12\catcode `\_12\catcode `\%12\relax}%
\providecommand \@@startlink[1]{}%
\providecommand \@@endlink[0]{}%
\providecommand \url  [0]{\begingroup\@sanitize@url \@url }%
\providecommand \@url [1]{\endgroup\@href {#1}{\urlprefix }}%
\providecommand \urlprefix  [0]{URL }%
\providecommand \Eprint [0]{\href }%
\providecommand \doibase [0]{https://doi.org/}%
\providecommand \selectlanguage [0]{\@gobble}%
\providecommand \bibinfo  [0]{\@secondoftwo}%
\providecommand \bibfield  [0]{\@secondoftwo}%
\providecommand \translation [1]{[#1]}%
\providecommand \BibitemOpen [0]{}%
\providecommand \bibitemStop [0]{}%
\providecommand \bibitemNoStop [0]{.\EOS\space}%
\providecommand \EOS [0]{\spacefactor3000\relax}%
\providecommand \BibitemShut  [1]{\csname bibitem#1\endcsname}%
\let\auto@bib@innerbib\@empty
\bibitem [{\citenamefont {Laflorencie}(2016)}]{Laflorencie16}%
  \BibitemOpen
  \bibfield  {author} {\bibinfo {author} {\bibfnamefont {N.}~\bibnamefont
  {Laflorencie}},\ }\href
  {https://doi.org/http://dx.doi.org/10.1016/j.physrep.2016.06.008} {\bibfield
  {journal} {\bibinfo  {journal} {Physics Reports}\ }\textbf {\bibinfo {volume}
  {646}},\ \bibinfo {pages} {1 } (\bibinfo {year} {2016})},\ \bibinfo {note}
  {quantum entanglement in condensed matter systems}\BibitemShut {NoStop}%
\bibitem [{\citenamefont {Calabrese}\ \emph {et~al.}(2009)\citenamefont
  {Calabrese}, \citenamefont {Cardy},\ and\ \citenamefont
  {Doyon}}]{Calabrese09_ex}%
  \BibitemOpen
  \bibfield  {author} {\bibinfo {author} {\bibfnamefont {P.}~\bibnamefont
  {Calabrese}}, \bibinfo {author} {\bibfnamefont {J.}~\bibnamefont {Cardy}},\
  and\ \bibinfo {author} {\bibfnamefont {B.}~\bibnamefont {Doyon}},\ }\href
  {https://doi.org/10.1088/1751-8121/42/50/500301} {\bibfield  {journal}
  {\bibinfo  {journal} {Journal of Physics A: Mathematical and Theoretical}\
  }\textbf {\bibinfo {volume} {42}},\ \bibinfo {pages} {500301} (\bibinfo
  {year} {2009})}\BibitemShut {NoStop}%
\bibitem [{\citenamefont {Amico}\ \emph {et~al.}(2008)\citenamefont {Amico},
  \citenamefont {Fazio}, \citenamefont {Osterloh},\ and\ \citenamefont
  {Vedral}}]{Amico08}%
  \BibitemOpen
  \bibfield  {author} {\bibinfo {author} {\bibfnamefont {L.}~\bibnamefont
  {Amico}}, \bibinfo {author} {\bibfnamefont {R.}~\bibnamefont {Fazio}},
  \bibinfo {author} {\bibfnamefont {A.}~\bibnamefont {Osterloh}},\ and\
  \bibinfo {author} {\bibfnamefont {V.}~\bibnamefont {Vedral}},\ }\href
  {https://doi.org/10.1103/RevModPhys.80.517} {\bibfield  {journal} {\bibinfo
  {journal} {Rev. Mod. Phys.}\ }\textbf {\bibinfo {volume} {80}},\ \bibinfo
  {pages} {517} (\bibinfo {year} {2008})}\BibitemShut {NoStop}%
\bibitem [{\citenamefont {Srednicki}(1993)}]{Srednicki93}%
  \BibitemOpen
  \bibfield  {author} {\bibinfo {author} {\bibfnamefont {M.}~\bibnamefont
  {Srednicki}},\ }\href {https://doi.org/10.1103/PhysRevLett.71.666} {\bibfield
   {journal} {\bibinfo  {journal} {Phys. Rev. Lett.}\ }\textbf {\bibinfo
  {volume} {71}},\ \bibinfo {pages} {666} (\bibinfo {year} {1993})}\BibitemShut
  {NoStop}%
\bibitem [{\citenamefont {Eisert}\ \emph {et~al.}(2010)\citenamefont {Eisert},
  \citenamefont {Cramer},\ and\ \citenamefont {Plenio}}]{Eisert10}%
  \BibitemOpen
  \bibfield  {author} {\bibinfo {author} {\bibfnamefont {J.}~\bibnamefont
  {Eisert}}, \bibinfo {author} {\bibfnamefont {M.}~\bibnamefont {Cramer}},\
  and\ \bibinfo {author} {\bibfnamefont {M.~B.}\ \bibnamefont {Plenio}},\
  }\href {https://doi.org/10.1103/RevModPhys.82.277} {\bibfield  {journal}
  {\bibinfo  {journal} {Rev. Mod. Phys.}\ }\textbf {\bibinfo {volume} {82}},\
  \bibinfo {pages} {277} (\bibinfo {year} {2010})}\BibitemShut {NoStop}%
\bibitem [{\citenamefont {Holzhey}\ \emph {et~al.}(1994)\citenamefont
  {Holzhey}, \citenamefont {Larsen},\ and\ \citenamefont
  {Wilczek}}]{Holzhey94}%
  \BibitemOpen
  \bibfield  {author} {\bibinfo {author} {\bibfnamefont {C.}~\bibnamefont
  {Holzhey}}, \bibinfo {author} {\bibfnamefont {F.}~\bibnamefont {Larsen}},\
  and\ \bibinfo {author} {\bibfnamefont {F.}~\bibnamefont {Wilczek}},\ }\href
  {https://doi.org/http://dx.doi.org/10.1016/0550-3213(94)90402-2} {\bibfield
  {journal} {\bibinfo  {journal} {Nuclear Physics B}\ }\textbf {\bibinfo
  {volume} {424}},\ \bibinfo {pages} {443 } (\bibinfo {year}
  {1994})}\BibitemShut {NoStop}%
\bibitem [{\citenamefont {Vidal}\ \emph {et~al.}(2003)\citenamefont {Vidal},
  \citenamefont {Latorre}, \citenamefont {Rico},\ and\ \citenamefont
  {Kitaev}}]{Vidal03}%
  \BibitemOpen
  \bibfield  {author} {\bibinfo {author} {\bibfnamefont {G.}~\bibnamefont
  {Vidal}}, \bibinfo {author} {\bibfnamefont {J.~I.}\ \bibnamefont {Latorre}},
  \bibinfo {author} {\bibfnamefont {E.}~\bibnamefont {Rico}},\ and\ \bibinfo
  {author} {\bibfnamefont {A.}~\bibnamefont {Kitaev}},\ }\href
  {https://doi.org/10.1103/PhysRevLett.90.227902} {\bibfield  {journal}
  {\bibinfo  {journal} {Phys. Rev. Lett.}\ }\textbf {\bibinfo {volume} {90}},\
  \bibinfo {pages} {227902} (\bibinfo {year} {2003})}\BibitemShut {NoStop}%
\bibitem [{\citenamefont {Calabrese}\ and\ \citenamefont
  {Cardy}(2004)}]{Calabrese04}%
  \BibitemOpen
  \bibfield  {author} {\bibinfo {author} {\bibfnamefont {P.}~\bibnamefont
  {Calabrese}}\ and\ \bibinfo {author} {\bibfnamefont {J.}~\bibnamefont
  {Cardy}},\ }\href {http://stacks.iop.org/1742-5468/2004/i=06/a=P06002}
  {\bibfield  {journal} {\bibinfo  {journal} {Journal of Statistical Mechanics:
  Theory and Experiment}\ }\textbf {\bibinfo {volume} {2004}},\ \bibinfo
  {pages} {P06002} (\bibinfo {year} {2004})}\BibitemShut {NoStop}%
\bibitem [{\citenamefont {Calabrese}\ and\ \citenamefont
  {Cardy}(2009)}]{Calabrese09}%
  \BibitemOpen
  \bibfield  {author} {\bibinfo {author} {\bibfnamefont {P.}~\bibnamefont
  {Calabrese}}\ and\ \bibinfo {author} {\bibfnamefont {J.}~\bibnamefont
  {Cardy}},\ }\href {https://doi.org/10.1088/1751-8113/42/50/504005} {\bibfield
   {journal} {\bibinfo  {journal} {Journal of Physics A: Mathematical and
  Theoretical}\ }\textbf {\bibinfo {volume} {42}},\ \bibinfo {pages} {504005}
  (\bibinfo {year} {2009})}\BibitemShut {NoStop}%
\bibitem [{\citenamefont {Kitaev}\ and\ \citenamefont
  {Preskill}(2006)}]{Kitaev06}%
  \BibitemOpen
  \bibfield  {author} {\bibinfo {author} {\bibfnamefont {A.}~\bibnamefont
  {Kitaev}}\ and\ \bibinfo {author} {\bibfnamefont {J.}~\bibnamefont
  {Preskill}},\ }\href {https://doi.org/10.1103/PhysRevLett.96.110404}
  {\bibfield  {journal} {\bibinfo  {journal} {Phys. Rev. Lett.}\ }\textbf
  {\bibinfo {volume} {96}},\ \bibinfo {pages} {110404} (\bibinfo {year}
  {2006})}\BibitemShut {NoStop}%
\bibitem [{\citenamefont {Levin}\ and\ \citenamefont {Wen}(2006)}]{Levin06}%
  \BibitemOpen
  \bibfield  {author} {\bibinfo {author} {\bibfnamefont {M.}~\bibnamefont
  {Levin}}\ and\ \bibinfo {author} {\bibfnamefont {X.-G.}\ \bibnamefont
  {Wen}},\ }\href {https://doi.org/10.1103/PhysRevLett.96.110405} {\bibfield
  {journal} {\bibinfo  {journal} {Phys. Rev. Lett.}\ }\textbf {\bibinfo
  {volume} {96}},\ \bibinfo {pages} {110405} (\bibinfo {year}
  {2006})}\BibitemShut {NoStop}%
\bibitem [{\citenamefont {Hamma}\ \emph
  {et~al.}(2005{\natexlab{a}})\citenamefont {Hamma}, \citenamefont
  {Ionicioiu},\ and\ \citenamefont {Zanardi}}]{Hamma05}%
  \BibitemOpen
  \bibfield  {author} {\bibinfo {author} {\bibfnamefont {A.}~\bibnamefont
  {Hamma}}, \bibinfo {author} {\bibfnamefont {R.}~\bibnamefont {Ionicioiu}},\
  and\ \bibinfo {author} {\bibfnamefont {P.}~\bibnamefont {Zanardi}},\ }\href
  {https://doi.org/10.1103/PhysRevA.71.022315} {\bibfield  {journal} {\bibinfo
  {journal} {Phys. Rev. A}\ }\textbf {\bibinfo {volume} {71}},\ \bibinfo
  {pages} {022315} (\bibinfo {year} {2005}{\natexlab{a}})}\BibitemShut
  {NoStop}%
\bibitem [{\citenamefont {Hamma}\ \emph
  {et~al.}(2005{\natexlab{b}})\citenamefont {Hamma}, \citenamefont
  {Ionicioiu},\ and\ \citenamefont {Zanardi}}]{Hamma05PhysLettA}%
  \BibitemOpen
  \bibfield  {author} {\bibinfo {author} {\bibfnamefont {A.}~\bibnamefont
  {Hamma}}, \bibinfo {author} {\bibfnamefont {R.}~\bibnamefont {Ionicioiu}},\
  and\ \bibinfo {author} {\bibfnamefont {P.}~\bibnamefont {Zanardi}},\ }\href
  {https://doi.org/http://dx.doi.org/10.1016/j.physleta.2005.01.060} {\bibfield
   {journal} {\bibinfo  {journal} {Physics Letters A}\ }\textbf {\bibinfo
  {volume} {337}},\ \bibinfo {pages} {22 } (\bibinfo {year}
  {2005}{\natexlab{b}})}\BibitemShut {NoStop}%
\bibitem [{\citenamefont {Li}\ and\ \citenamefont {Haldane}(2008)}]{Li08}%
  \BibitemOpen
  \bibfield  {author} {\bibinfo {author} {\bibfnamefont {H.}~\bibnamefont
  {Li}}\ and\ \bibinfo {author} {\bibfnamefont {F.~D.~M.}\ \bibnamefont
  {Haldane}},\ }\href {https://doi.org/10.1103/PhysRevLett.101.010504}
  {\bibfield  {journal} {\bibinfo  {journal} {Phys. Rev. Lett.}\ }\textbf
  {\bibinfo {volume} {101}},\ \bibinfo {pages} {010504} (\bibinfo {year}
  {2008})}\BibitemShut {NoStop}%
\bibitem [{\citenamefont {Thomale}\ \emph {et~al.}(2010)\citenamefont
  {Thomale}, \citenamefont {Sterdyniak}, \citenamefont {Regnault},\ and\
  \citenamefont {Bernevig}}]{Thomale10FQH}%
  \BibitemOpen
  \bibfield  {author} {\bibinfo {author} {\bibfnamefont {R.}~\bibnamefont
  {Thomale}}, \bibinfo {author} {\bibfnamefont {A.}~\bibnamefont {Sterdyniak}},
  \bibinfo {author} {\bibfnamefont {N.}~\bibnamefont {Regnault}},\ and\
  \bibinfo {author} {\bibfnamefont {B.~A.}\ \bibnamefont {Bernevig}},\ }\href
  {https://doi.org/10.1103/PhysRevLett.104.180502} {\bibfield  {journal}
  {\bibinfo  {journal} {Phys. Rev. Lett.}\ }\textbf {\bibinfo {volume} {104}},\
  \bibinfo {pages} {180502} (\bibinfo {year} {2010})}\BibitemShut {NoStop}%
\bibitem [{\citenamefont {Sterdyniak}\ \emph {et~al.}(2012)\citenamefont
  {Sterdyniak}, \citenamefont {Chandran}, \citenamefont {Regnault},
  \citenamefont {Bernevig},\ and\ \citenamefont {Bonderson}}]{Sterdyniak12}%
  \BibitemOpen
  \bibfield  {author} {\bibinfo {author} {\bibfnamefont {A.}~\bibnamefont
  {Sterdyniak}}, \bibinfo {author} {\bibfnamefont {A.}~\bibnamefont
  {Chandran}}, \bibinfo {author} {\bibfnamefont {N.}~\bibnamefont {Regnault}},
  \bibinfo {author} {\bibfnamefont {B.~A.}\ \bibnamefont {Bernevig}},\ and\
  \bibinfo {author} {\bibfnamefont {P.}~\bibnamefont {Bonderson}},\ }\href
  {https://doi.org/10.1103/PhysRevB.85.125308} {\bibfield  {journal} {\bibinfo
  {journal} {Phys. Rev. B}\ }\textbf {\bibinfo {volume} {85}},\ \bibinfo
  {pages} {125308} (\bibinfo {year} {2012})}\BibitemShut {NoStop}%
\bibitem [{\citenamefont {Dubail}\ \emph
  {et~al.}(2012{\natexlab{a}})\citenamefont {Dubail}, \citenamefont {Read},\
  and\ \citenamefont {Rezayi}}]{Dubail12FQH}%
  \BibitemOpen
  \bibfield  {author} {\bibinfo {author} {\bibfnamefont {J.}~\bibnamefont
  {Dubail}}, \bibinfo {author} {\bibfnamefont {N.}~\bibnamefont {Read}},\ and\
  \bibinfo {author} {\bibfnamefont {E.~H.}\ \bibnamefont {Rezayi}},\ }\href
  {https://doi.org/10.1103/PhysRevB.85.115321} {\bibfield  {journal} {\bibinfo
  {journal} {Phys. Rev. B}\ }\textbf {\bibinfo {volume} {85}},\ \bibinfo
  {pages} {115321} (\bibinfo {year} {2012}{\natexlab{a}})}\BibitemShut
  {NoStop}%
\bibitem [{\citenamefont {Rodriguez}\ \emph {et~al.}(2012)\citenamefont
  {Rodriguez}, \citenamefont {Simon},\ and\ \citenamefont
  {Slingerland}}]{Rodriguez12}%
  \BibitemOpen
  \bibfield  {author} {\bibinfo {author} {\bibfnamefont {I.~D.}\ \bibnamefont
  {Rodriguez}}, \bibinfo {author} {\bibfnamefont {S.~H.}\ \bibnamefont
  {Simon}},\ and\ \bibinfo {author} {\bibfnamefont {J.~K.}\ \bibnamefont
  {Slingerland}},\ }\href {https://doi.org/10.1103/PhysRevLett.108.256806}
  {\bibfield  {journal} {\bibinfo  {journal} {Phys. Rev. Lett.}\ }\textbf
  {\bibinfo {volume} {108}},\ \bibinfo {pages} {256806} (\bibinfo {year}
  {2012})}\BibitemShut {NoStop}%
\bibitem [{\citenamefont {Rodriguez}\ \emph {et~al.}(2013)\citenamefont
  {Rodriguez}, \citenamefont {Davenport}, \citenamefont {Simon},\ and\
  \citenamefont {Slingerland}}]{Rodriguez13}%
  \BibitemOpen
  \bibfield  {author} {\bibinfo {author} {\bibfnamefont {I.~D.}\ \bibnamefont
  {Rodriguez}}, \bibinfo {author} {\bibfnamefont {S.~C.}\ \bibnamefont
  {Davenport}}, \bibinfo {author} {\bibfnamefont {S.~H.}\ \bibnamefont
  {Simon}},\ and\ \bibinfo {author} {\bibfnamefont {J.~K.}\ \bibnamefont
  {Slingerland}},\ }\href {https://doi.org/10.1103/PhysRevB.88.155307}
  {\bibfield  {journal} {\bibinfo  {journal} {Phys. Rev. B}\ }\textbf {\bibinfo
  {volume} {88}},\ \bibinfo {pages} {155307} (\bibinfo {year}
  {2013})}\BibitemShut {NoStop}%
\bibitem [{\citenamefont {Yan}\ \emph {et~al.}(2019)\citenamefont {Yan},
  \citenamefont {Biswas},\ and\ \citenamefont {Greene}}]{Yan19}%
  \BibitemOpen
  \bibfield  {author} {\bibinfo {author} {\bibfnamefont {B.}~\bibnamefont
  {Yan}}, \bibinfo {author} {\bibfnamefont {R.~R.}\ \bibnamefont {Biswas}},\
  and\ \bibinfo {author} {\bibfnamefont {C.~H.}\ \bibnamefont {Greene}},\
  }\href {https://doi.org/10.1103/PhysRevB.99.035153} {\bibfield  {journal}
  {\bibinfo  {journal} {Phys. Rev. B}\ }\textbf {\bibinfo {volume} {99}},\
  \bibinfo {pages} {035153} (\bibinfo {year} {2019})}\BibitemShut {NoStop}%
\bibitem [{\citenamefont {Regnault}\ and\ \citenamefont
  {Bernevig}(2011)}]{Regnault11}%
  \BibitemOpen
  \bibfield  {author} {\bibinfo {author} {\bibfnamefont {N.}~\bibnamefont
  {Regnault}}\ and\ \bibinfo {author} {\bibfnamefont {B.~A.}\ \bibnamefont
  {Bernevig}},\ }\href {https://doi.org/10.1103/PhysRevX.1.021014} {\bibfield
  {journal} {\bibinfo  {journal} {Phys. Rev. X}\ }\textbf {\bibinfo {volume}
  {1}},\ \bibinfo {pages} {021014} (\bibinfo {year} {2011})}\BibitemShut
  {NoStop}%
\bibitem [{\citenamefont {Turner}\ \emph {et~al.}(2010)\citenamefont {Turner},
  \citenamefont {Zhang},\ and\ \citenamefont {Vishwanath}}]{Turner10}%
  \BibitemOpen
  \bibfield  {author} {\bibinfo {author} {\bibfnamefont {A.~M.}\ \bibnamefont
  {Turner}}, \bibinfo {author} {\bibfnamefont {Y.}~\bibnamefont {Zhang}},\ and\
  \bibinfo {author} {\bibfnamefont {A.}~\bibnamefont {Vishwanath}},\ }\href
  {https://doi.org/10.1103/PhysRevB.82.241102} {\bibfield  {journal} {\bibinfo
  {journal} {Phys. Rev. B}\ }\textbf {\bibinfo {volume} {82}},\ \bibinfo
  {pages} {241102(R)} (\bibinfo {year} {2010})}\BibitemShut {NoStop}%
\bibitem [{\citenamefont {Fidkowski}(2010)}]{Fidkowski10}%
  \BibitemOpen
  \bibfield  {author} {\bibinfo {author} {\bibfnamefont {L.}~\bibnamefont
  {Fidkowski}},\ }\href {https://doi.org/10.1103/PhysRevLett.104.130502}
  {\bibfield  {journal} {\bibinfo  {journal} {Phys. Rev. Lett.}\ }\textbf
  {\bibinfo {volume} {104}},\ \bibinfo {pages} {130502} (\bibinfo {year}
  {2010})}\BibitemShut {NoStop}%
\bibitem [{\citenamefont {Alexandradinata}\ \emph {et~al.}(2011)\citenamefont
  {Alexandradinata}, \citenamefont {Hughes},\ and\ \citenamefont
  {Bernevig}}]{Alexandradinata11}%
  \BibitemOpen
  \bibfield  {author} {\bibinfo {author} {\bibfnamefont {A.}~\bibnamefont
  {Alexandradinata}}, \bibinfo {author} {\bibfnamefont {T.~L.}\ \bibnamefont
  {Hughes}},\ and\ \bibinfo {author} {\bibfnamefont {B.~A.}\ \bibnamefont
  {Bernevig}},\ }\href {https://doi.org/10.1103/PhysRevB.84.195103} {\bibfield
  {journal} {\bibinfo  {journal} {Phys. Rev. B}\ }\textbf {\bibinfo {volume}
  {84}},\ \bibinfo {pages} {195103} (\bibinfo {year} {2011})}\BibitemShut
  {NoStop}%
\bibitem [{\citenamefont {Fang}\ \emph {et~al.}(2013)\citenamefont {Fang},
  \citenamefont {Gilbert},\ and\ \citenamefont {Bernevig}}]{Fang13}%
  \BibitemOpen
  \bibfield  {author} {\bibinfo {author} {\bibfnamefont {C.}~\bibnamefont
  {Fang}}, \bibinfo {author} {\bibfnamefont {M.~J.}\ \bibnamefont {Gilbert}},\
  and\ \bibinfo {author} {\bibfnamefont {B.~A.}\ \bibnamefont {Bernevig}},\
  }\href {https://doi.org/10.1103/PhysRevB.87.035119} {\bibfield  {journal}
  {\bibinfo  {journal} {Phys. Rev. B}\ }\textbf {\bibinfo {volume} {87}},\
  \bibinfo {pages} {035119} (\bibinfo {year} {2013})}\BibitemShut {NoStop}%
\bibitem [{\citenamefont {Pollmann}\ \emph {et~al.}(2010)\citenamefont
  {Pollmann}, \citenamefont {Turner}, \citenamefont {Berg},\ and\ \citenamefont
  {Oshikawa}}]{Pollmann10}%
  \BibitemOpen
  \bibfield  {author} {\bibinfo {author} {\bibfnamefont {F.}~\bibnamefont
  {Pollmann}}, \bibinfo {author} {\bibfnamefont {A.~M.}\ \bibnamefont
  {Turner}}, \bibinfo {author} {\bibfnamefont {E.}~\bibnamefont {Berg}},\ and\
  \bibinfo {author} {\bibfnamefont {M.}~\bibnamefont {Oshikawa}},\ }\href
  {https://doi.org/10.1103/PhysRevB.81.064439} {\bibfield  {journal} {\bibinfo
  {journal} {Phys. Rev. B}\ }\textbf {\bibinfo {volume} {81}},\ \bibinfo
  {pages} {064439} (\bibinfo {year} {2010})}\BibitemShut {NoStop}%
\bibitem [{\citenamefont {Turner}\ \emph {et~al.}(2011)\citenamefont {Turner},
  \citenamefont {Pollmann},\ and\ \citenamefont {Berg}}]{Turner11}%
  \BibitemOpen
  \bibfield  {author} {\bibinfo {author} {\bibfnamefont {A.~M.}\ \bibnamefont
  {Turner}}, \bibinfo {author} {\bibfnamefont {F.}~\bibnamefont {Pollmann}},\
  and\ \bibinfo {author} {\bibfnamefont {E.}~\bibnamefont {Berg}},\ }\href
  {https://doi.org/10.1103/PhysRevB.83.075102} {\bibfield  {journal} {\bibinfo
  {journal} {Phys. Rev. B}\ }\textbf {\bibinfo {volume} {83}},\ \bibinfo
  {pages} {075102} (\bibinfo {year} {2011})}\BibitemShut {NoStop}%
\bibitem [{\citenamefont {Fidkowski}\ and\ \citenamefont
  {Kitaev}(2011)}]{Fidkowski11}%
  \BibitemOpen
  \bibfield  {author} {\bibinfo {author} {\bibfnamefont {L.}~\bibnamefont
  {Fidkowski}}\ and\ \bibinfo {author} {\bibfnamefont {A.}~\bibnamefont
  {Kitaev}},\ }\href {https://doi.org/10.1103/PhysRevB.83.075103} {\bibfield
  {journal} {\bibinfo  {journal} {Phys. Rev. B}\ }\textbf {\bibinfo {volume}
  {83}},\ \bibinfo {pages} {075103} (\bibinfo {year} {2011})}\BibitemShut
  {NoStop}%
\bibitem [{\citenamefont {Cho}\ \emph {et~al.}(2017)\citenamefont {Cho},
  \citenamefont {Shiozaki}, \citenamefont {Ryu},\ and\ \citenamefont
  {Ludwig}}]{Cho17}%
  \BibitemOpen
  \bibfield  {author} {\bibinfo {author} {\bibfnamefont {G.~Y.}\ \bibnamefont
  {Cho}}, \bibinfo {author} {\bibfnamefont {K.}~\bibnamefont {Shiozaki}},
  \bibinfo {author} {\bibfnamefont {S.}~\bibnamefont {Ryu}},\ and\ \bibinfo
  {author} {\bibfnamefont {A.~W.~W.}\ \bibnamefont {Ludwig}},\ }\href
  {https://doi.org/10.1088/1751-8121/aa7782} {\bibfield  {journal} {\bibinfo
  {journal} {Journal of Physics A: Mathematical and Theoretical}\ }\textbf
  {\bibinfo {volume} {50}},\ \bibinfo {pages} {304002} (\bibinfo {year}
  {2017})}\BibitemShut {NoStop}%
\bibitem [{\citenamefont {Qi}\ \emph {et~al.}(2012)\citenamefont {Qi},
  \citenamefont {Katsura},\ and\ \citenamefont {Ludwig}}]{Qi12}%
  \BibitemOpen
  \bibfield  {author} {\bibinfo {author} {\bibfnamefont {X.-L.}\ \bibnamefont
  {Qi}}, \bibinfo {author} {\bibfnamefont {H.}~\bibnamefont {Katsura}},\ and\
  \bibinfo {author} {\bibfnamefont {A.~W.~W.}\ \bibnamefont {Ludwig}},\ }\href
  {https://doi.org/10.1103/PhysRevLett.108.196402} {\bibfield  {journal}
  {\bibinfo  {journal} {Phys. Rev. Lett.}\ }\textbf {\bibinfo {volume} {108}},\
  \bibinfo {pages} {196402} (\bibinfo {year} {2012})}\BibitemShut {NoStop}%
\bibitem [{\citenamefont {Chandran}\ \emph {et~al.}(2011)\citenamefont
  {Chandran}, \citenamefont {Hermanns}, \citenamefont {Regnault},\ and\
  \citenamefont {Bernevig}}]{Chandran11}%
  \BibitemOpen
  \bibfield  {author} {\bibinfo {author} {\bibfnamefont {A.}~\bibnamefont
  {Chandran}}, \bibinfo {author} {\bibfnamefont {M.}~\bibnamefont {Hermanns}},
  \bibinfo {author} {\bibfnamefont {N.}~\bibnamefont {Regnault}},\ and\
  \bibinfo {author} {\bibfnamefont {B.~A.}\ \bibnamefont {Bernevig}},\ }\href
  {https://doi.org/10.1103/PhysRevB.84.205136} {\bibfield  {journal} {\bibinfo
  {journal} {Phys. Rev. B}\ }\textbf {\bibinfo {volume} {84}},\ \bibinfo
  {pages} {205136} (\bibinfo {year} {2011})}\BibitemShut {NoStop}%
\bibitem [{\citenamefont {Dubail}\ \emph
  {et~al.}(2012{\natexlab{b}})\citenamefont {Dubail}, \citenamefont {Read},\
  and\ \citenamefont {Rezayi}}]{Dubail12proof}%
  \BibitemOpen
  \bibfield  {author} {\bibinfo {author} {\bibfnamefont {J.}~\bibnamefont
  {Dubail}}, \bibinfo {author} {\bibfnamefont {N.}~\bibnamefont {Read}},\ and\
  \bibinfo {author} {\bibfnamefont {E.~H.}\ \bibnamefont {Rezayi}},\ }\href
  {https://doi.org/10.1103/PhysRevB.86.245310} {\bibfield  {journal} {\bibinfo
  {journal} {Phys. Rev. B}\ }\textbf {\bibinfo {volume} {86}},\ \bibinfo
  {pages} {245310} (\bibinfo {year} {2012}{\natexlab{b}})}\BibitemShut
  {NoStop}%
\bibitem [{\citenamefont {Swingle}\ and\ \citenamefont
  {Senthil}(2012)}]{SwingleSenthil12}%
  \BibitemOpen
  \bibfield  {author} {\bibinfo {author} {\bibfnamefont {B.}~\bibnamefont
  {Swingle}}\ and\ \bibinfo {author} {\bibfnamefont {T.}~\bibnamefont
  {Senthil}},\ }\href {https://doi.org/10.1103/PhysRevB.86.045117} {\bibfield
  {journal} {\bibinfo  {journal} {Phys. Rev. B}\ }\textbf {\bibinfo {volume}
  {86}},\ \bibinfo {pages} {045117} (\bibinfo {year} {2012})}\BibitemShut
  {NoStop}%
\bibitem [{\citenamefont {Lundgren}\ \emph {et~al.}(2013)\citenamefont
  {Lundgren}, \citenamefont {Fuji}, \citenamefont {Furukawa},\ and\
  \citenamefont {Oshikawa}}]{Lundgren13}%
  \BibitemOpen
  \bibfield  {author} {\bibinfo {author} {\bibfnamefont {R.}~\bibnamefont
  {Lundgren}}, \bibinfo {author} {\bibfnamefont {Y.}~\bibnamefont {Fuji}},
  \bibinfo {author} {\bibfnamefont {S.}~\bibnamefont {Furukawa}},\ and\
  \bibinfo {author} {\bibfnamefont {M.}~\bibnamefont {Oshikawa}},\ }\href
  {https://doi.org/10.1103/PhysRevB.88.245137} {\bibfield  {journal} {\bibinfo
  {journal} {Phys. Rev. B}\ }\textbf {\bibinfo {volume} {88}},\ \bibinfo
  {pages} {245137} (\bibinfo {year} {2013})}\BibitemShut {NoStop}%
\bibitem [{\citenamefont {Cano}\ \emph {et~al.}(2015)\citenamefont {Cano},
  \citenamefont {Hughes},\ and\ \citenamefont {Mulligan}}]{Cano15}%
  \BibitemOpen
  \bibfield  {author} {\bibinfo {author} {\bibfnamefont {J.}~\bibnamefont
  {Cano}}, \bibinfo {author} {\bibfnamefont {T.~L.}\ \bibnamefont {Hughes}},\
  and\ \bibinfo {author} {\bibfnamefont {M.}~\bibnamefont {Mulligan}},\ }\href
  {https://doi.org/10.1103/PhysRevB.92.075104} {\bibfield  {journal} {\bibinfo
  {journal} {Phys. Rev. B}\ }\textbf {\bibinfo {volume} {92}},\ \bibinfo
  {pages} {075104} (\bibinfo {year} {2015})}\BibitemShut {NoStop}%
\bibitem [{\citenamefont {Das}\ and\ \citenamefont {Datta}(2015)}]{Das15}%
  \BibitemOpen
  \bibfield  {author} {\bibinfo {author} {\bibfnamefont {D.}~\bibnamefont
  {Das}}\ and\ \bibinfo {author} {\bibfnamefont {S.}~\bibnamefont {Datta}},\
  }\href {https://doi.org/10.1103/PhysRevLett.115.131602} {\bibfield  {journal}
  {\bibinfo  {journal} {Phys. Rev. Lett.}\ }\textbf {\bibinfo {volume} {115}},\
  \bibinfo {pages} {131602} (\bibinfo {year} {2015})}\BibitemShut {NoStop}%
\bibitem [{\citenamefont {Wen}\ \emph {et~al.}(2016)\citenamefont {Wen},
  \citenamefont {Matsuura},\ and\ \citenamefont {Ryu}}]{WenX16}%
  \BibitemOpen
  \bibfield  {author} {\bibinfo {author} {\bibfnamefont {X.}~\bibnamefont
  {Wen}}, \bibinfo {author} {\bibfnamefont {S.}~\bibnamefont {Matsuura}},\ and\
  \bibinfo {author} {\bibfnamefont {S.}~\bibnamefont {Ryu}},\ }\href
  {https://doi.org/10.1103/PhysRevB.93.245140} {\bibfield  {journal} {\bibinfo
  {journal} {Phys. Rev. B}\ }\textbf {\bibinfo {volume} {93}},\ \bibinfo
  {pages} {245140} (\bibinfo {year} {2016})}\BibitemShut {NoStop}%
\bibitem [{\citenamefont {Sohal}\ \emph {et~al.}(2020)\citenamefont {Sohal},
  \citenamefont {Han}, \citenamefont {Santos},\ and\ \citenamefont
  {Teo}}]{Sohal20}%
  \BibitemOpen
  \bibfield  {author} {\bibinfo {author} {\bibfnamefont {R.}~\bibnamefont
  {Sohal}}, \bibinfo {author} {\bibfnamefont {B.}~\bibnamefont {Han}}, \bibinfo
  {author} {\bibfnamefont {L.~H.}\ \bibnamefont {Santos}},\ and\ \bibinfo
  {author} {\bibfnamefont {J.~C.~Y.}\ \bibnamefont {Teo}},\ }\href
  {https://doi.org/10.1103/PhysRevB.102.045102} {\bibfield  {journal} {\bibinfo
   {journal} {Phys. Rev. B}\ }\textbf {\bibinfo {volume} {102}},\ \bibinfo
  {pages} {045102} (\bibinfo {year} {2020})}\BibitemShut {NoStop}%
\bibitem [{\citenamefont {Poilblanc}(2010)}]{Poilblanc10}%
  \BibitemOpen
  \bibfield  {author} {\bibinfo {author} {\bibfnamefont {D.}~\bibnamefont
  {Poilblanc}},\ }\href {https://doi.org/10.1103/PhysRevLett.105.077202}
  {\bibfield  {journal} {\bibinfo  {journal} {Phys. Rev. Lett.}\ }\textbf
  {\bibinfo {volume} {105}},\ \bibinfo {pages} {077202} (\bibinfo {year}
  {2010})}\BibitemShut {NoStop}%
\bibitem [{\citenamefont {{Peschel}}\ and\ \citenamefont
  {{Chung}}(2011)}]{Peschel11}%
  \BibitemOpen
  \bibfield  {author} {\bibinfo {author} {\bibfnamefont {I.}~\bibnamefont
  {{Peschel}}}\ and\ \bibinfo {author} {\bibfnamefont {M.-C.}\ \bibnamefont
  {{Chung}}},\ }\href {https://doi.org/10.1209/0295-5075/96/50006} {\bibfield
  {journal} {\bibinfo  {journal} {EPL (Europhysics Letters)}\ }\textbf
  {\bibinfo {volume} {96}},\ \bibinfo {pages} {50006} (\bibinfo {year}
  {2011})}\BibitemShut {NoStop}%
\bibitem [{\citenamefont {L\"auchli}\ and\ \citenamefont
  {Schliemann}(2012)}]{Lauchli12}%
  \BibitemOpen
  \bibfield  {author} {\bibinfo {author} {\bibfnamefont {A.~M.}\ \bibnamefont
  {L\"auchli}}\ and\ \bibinfo {author} {\bibfnamefont {J.}~\bibnamefont
  {Schliemann}},\ }\href@noop {} {\bibfield  {journal} {\bibinfo  {journal}
  {Phys. Rev. B}\ }\textbf {\bibinfo {volume} {85}},\ \bibinfo {pages} {054403}
  (\bibinfo {year} {2012})}\BibitemShut {NoStop}%
\bibitem [{\citenamefont {{Schliemann}}\ and\ \citenamefont
  {{L{\"a}uchli}}(2012)}]{Schliemann12}%
  \BibitemOpen
  \bibfield  {author} {\bibinfo {author} {\bibfnamefont {J.}~\bibnamefont
  {{Schliemann}}}\ and\ \bibinfo {author} {\bibfnamefont {A.~M.}\ \bibnamefont
  {{L{\"a}uchli}}},\ }\href {https://doi.org/10.1088/1742-5468/2012/11/P11021}
  {\bibfield  {journal} {\bibinfo  {journal} {J. Stat. Mech. Theor. Exp.}\
  }\textbf {\bibinfo {volume} {11}},\ \bibinfo {pages} {21} (\bibinfo {year}
  {2012})}\BibitemShut {NoStop}%
\bibitem [{\citenamefont {Furukawa}\ and\ \citenamefont
  {Kim}(2011)}]{Furukawa11}%
  \BibitemOpen
  \bibfield  {author} {\bibinfo {author} {\bibfnamefont {S.}~\bibnamefont
  {Furukawa}}\ and\ \bibinfo {author} {\bibfnamefont {Y.~B.}\ \bibnamefont
  {Kim}},\ }\href {https://doi.org/10.1103/PhysRevB.83.085112} {\bibfield
  {journal} {\bibinfo  {journal} {Phys. Rev. B}\ }\textbf {\bibinfo {volume}
  {83}},\ \bibinfo {pages} {085112} (\bibinfo {year} {2011})}\BibitemShut
  {NoStop}%
\bibitem [{\citenamefont {Chen}\ and\ \citenamefont {Fradkin}(2013)}]{Chen13}%
  \BibitemOpen
  \bibfield  {author} {\bibinfo {author} {\bibfnamefont {X.}~\bibnamefont
  {Chen}}\ and\ \bibinfo {author} {\bibfnamefont {E.}~\bibnamefont {Fradkin}},\
  }\href {http://stacks.iop.org/1742-5468/2013/i=08/a=P08013} {\bibfield
  {journal} {\bibinfo  {journal} {J. Stat. Mech. Theor. Exp.}\ }\textbf
  {\bibinfo {volume} {2013}},\ \bibinfo {pages} {P08013} (\bibinfo {year}
  {2013})}\BibitemShut {NoStop}%
\bibitem [{\citenamefont {Tanaka}\ \emph {et~al.}(2012)\citenamefont {Tanaka},
  \citenamefont {Tamura},\ and\ \citenamefont {Katsura}}]{Tanaka12}%
  \BibitemOpen
  \bibfield  {author} {\bibinfo {author} {\bibfnamefont {S.}~\bibnamefont
  {Tanaka}}, \bibinfo {author} {\bibfnamefont {R.}~\bibnamefont {Tamura}},\
  and\ \bibinfo {author} {\bibfnamefont {H.}~\bibnamefont {Katsura}},\ }\href
  {https://doi.org/10.1103/PhysRevA.86.032326} {\bibfield  {journal} {\bibinfo
  {journal} {Phys. Rev. A}\ }\textbf {\bibinfo {volume} {86}},\ \bibinfo
  {pages} {032326} (\bibinfo {year} {2012})}\BibitemShut {NoStop}%
\bibitem [{\citenamefont {Lundgren}\ \emph {et~al.}(2012)\citenamefont
  {Lundgren}, \citenamefont {Chua},\ and\ \citenamefont {Fiete}}]{Lundgren12}%
  \BibitemOpen
  \bibfield  {author} {\bibinfo {author} {\bibfnamefont {R.}~\bibnamefont
  {Lundgren}}, \bibinfo {author} {\bibfnamefont {V.}~\bibnamefont {Chua}},\
  and\ \bibinfo {author} {\bibfnamefont {G.~A.}\ \bibnamefont {Fiete}},\ }\href
  {https://doi.org/10.1103/PhysRevB.86.224422} {\bibfield  {journal} {\bibinfo
  {journal} {Phys. Rev. B}\ }\textbf {\bibinfo {volume} {86}},\ \bibinfo
  {pages} {224422} (\bibinfo {year} {2012})}\BibitemShut {NoStop}%
\bibitem [{\citenamefont {Santos}\ \emph {et~al.}(2016)\citenamefont {Santos},
  \citenamefont {Jian},\ and\ \citenamefont {Lundgren}}]{Santos16}%
  \BibitemOpen
  \bibfield  {author} {\bibinfo {author} {\bibfnamefont {R.~A.}\ \bibnamefont
  {Santos}}, \bibinfo {author} {\bibfnamefont {C.-M.}\ \bibnamefont {Jian}},\
  and\ \bibinfo {author} {\bibfnamefont {R.}~\bibnamefont {Lundgren}},\ }\href
  {https://doi.org/10.1103/PhysRevB.93.245101} {\bibfield  {journal} {\bibinfo
  {journal} {Phys. Rev. B}\ }\textbf {\bibinfo {volume} {93}},\ \bibinfo
  {pages} {245101} (\bibinfo {year} {2016})}\BibitemShut {NoStop}%
\bibitem [{\citenamefont {Fujita}\ \emph {et~al.}(2018)\citenamefont {Fujita},
  \citenamefont {Nakagawa}, \citenamefont {Sugiura},\ and\ \citenamefont
  {Oshikawa}}]{Fujita18}%
  \BibitemOpen
  \bibfield  {author} {\bibinfo {author} {\bibfnamefont {H.}~\bibnamefont
  {Fujita}}, \bibinfo {author} {\bibfnamefont {Y.~O.}\ \bibnamefont
  {Nakagawa}}, \bibinfo {author} {\bibfnamefont {S.}~\bibnamefont {Sugiura}},\
  and\ \bibinfo {author} {\bibfnamefont {M.}~\bibnamefont {Oshikawa}},\ }\href
  {https://doi.org/10.1103/PhysRevB.97.075114} {\bibfield  {journal} {\bibinfo
  {journal} {Phys. Rev. B}\ }\textbf {\bibinfo {volume} {97}},\ \bibinfo
  {pages} {075114} (\bibinfo {year} {2018})}\BibitemShut {NoStop}%
\bibitem [{\citenamefont {Schliemann}(2011)}]{Schliemann11}%
  \BibitemOpen
  \bibfield  {author} {\bibinfo {author} {\bibfnamefont {J.}~\bibnamefont
  {Schliemann}},\ }\href {https://doi.org/10.1103/PhysRevB.83.115322}
  {\bibfield  {journal} {\bibinfo  {journal} {Phys. Rev. B}\ }\textbf {\bibinfo
  {volume} {83}},\ \bibinfo {pages} {115322} (\bibinfo {year}
  {2011})}\BibitemShut {NoStop}%
\bibitem [{\citenamefont {{Schliemann}}(2013)}]{Schliemann13}%
  \BibitemOpen
  \bibfield  {author} {\bibinfo {author} {\bibfnamefont {J.}~\bibnamefont
  {{Schliemann}}},\ }\href {https://doi.org/10.1088/1367-2630/15/5/053017}
  {\bibfield  {journal} {\bibinfo  {journal} {New Journal of Physics}\ }\textbf
  {\bibinfo {volume} {15}},\ \bibinfo {eid} {053017} (\bibinfo {year}
  {2013})}\BibitemShut {NoStop}%
\bibitem [{\citenamefont {Schliemann}(2014)}]{Schliemann14}%
  \BibitemOpen
  \bibfield  {author} {\bibinfo {author} {\bibfnamefont {J.}~\bibnamefont
  {Schliemann}},\ }\href {https://doi.org/10.1088/1742-5468/2014/09/p09011}
  {\bibfield  {journal} {\bibinfo  {journal} {Journal of Statistical Mechanics:
  Theory and Experiment}\ }\textbf {\bibinfo {volume} {2014}},\ \bibinfo
  {pages} {P09011} (\bibinfo {year} {2014})}\BibitemShut {NoStop}%
\bibitem [{\citenamefont {{Ro{\'o}sz}}\ and\ \citenamefont
  {{Timm}}(2020)}]{Roosz20}%
  \BibitemOpen
  \bibfield  {author} {\bibinfo {author} {\bibfnamefont {G.}~\bibnamefont
  {{Ro{\'o}sz}}}\ and\ \bibinfo {author} {\bibfnamefont {C.}~\bibnamefont
  {{Timm}}},\ }\href@noop {} {\bibfield  {journal} {\bibinfo  {journal} {arXiv
  e-prints}\ ,\ \bibinfo {eid} {arXiv:2009.03274}} (\bibinfo {year} {2020})},\
  \Eprint {https://arxiv.org/abs/2009.03274} {arXiv:2009.03274
  [cond-mat.str-el]} \BibitemShut {NoStop}%
\bibitem [{\citenamefont {Xu}(2011)}]{Xu11}%
  \BibitemOpen
  \bibfield  {author} {\bibinfo {author} {\bibfnamefont {C.}~\bibnamefont
  {Xu}},\ }\href {https://doi.org/10.1103/PhysRevB.84.125119} {\bibfield
  {journal} {\bibinfo  {journal} {Phys. Rev. B}\ }\textbf {\bibinfo {volume}
  {84}},\ \bibinfo {pages} {125119} (\bibinfo {year} {2011})}\BibitemShut
  {NoStop}%
\bibitem [{\citenamefont {Mollabashi}\ \emph {et~al.}(2014)\citenamefont
  {Mollabashi}, \citenamefont {Shiba},\ and\ \citenamefont
  {Takayanagi}}]{Mollabashi14}%
  \BibitemOpen
  \bibfield  {author} {\bibinfo {author} {\bibfnamefont {A.}~\bibnamefont
  {Mollabashi}}, \bibinfo {author} {\bibfnamefont {N.}~\bibnamefont {Shiba}},\
  and\ \bibinfo {author} {\bibfnamefont {T.}~\bibnamefont {Takayanagi}},\
  }\href {https://doi.org/10.1007/JHEP04(2014)185} {\bibfield  {journal}
  {\bibinfo  {journal} {Journal of High Energy Physics}\ }\textbf {\bibinfo
  {volume} {2014}},\ \bibinfo {pages} {185} (\bibinfo {year}
  {2014})}\BibitemShut {NoStop}%
\bibitem [{\citenamefont {Mozaffar}(2016)}]{Mozaffar16}%
  \BibitemOpen
  \bibfield  {author} {\bibinfo {author} {\bibfnamefont {A.}~\bibnamefont
  {Mozaffar}, \bibfnamefont {M.~Reza Mohammadiand~Mollabashi}},\ }\href
  {https://doi.org/10.1007/JHEP03(2016)015} {\bibfield  {journal} {\bibinfo
  {journal} {Journal of High Energy Physics}\ }\textbf {\bibinfo {volume}
  {2016}},\ \bibinfo {pages} {15} (\bibinfo {year} {2016})}\BibitemShut
  {NoStop}%
\bibitem [{\citenamefont {Song}\ \emph {et~al.}(2011)\citenamefont {Song},
  \citenamefont {Laflorencie}, \citenamefont {Rachel},\ and\ \citenamefont
  {Le~Hur}}]{Song11}%
  \BibitemOpen
  \bibfield  {author} {\bibinfo {author} {\bibfnamefont {H.~F.}\ \bibnamefont
  {Song}}, \bibinfo {author} {\bibfnamefont {N.}~\bibnamefont {Laflorencie}},
  \bibinfo {author} {\bibfnamefont {S.}~\bibnamefont {Rachel}},\ and\ \bibinfo
  {author} {\bibfnamefont {K.}~\bibnamefont {Le~Hur}},\ }\href
  {https://doi.org/10.1103/PhysRevB.83.224410} {\bibfield  {journal} {\bibinfo
  {journal} {Phys. Rev. B}\ }\textbf {\bibinfo {volume} {83}},\ \bibinfo
  {pages} {224410} (\bibinfo {year} {2011})}\BibitemShut {NoStop}%
\bibitem [{\citenamefont {Kallin}\ \emph {et~al.}(2011)\citenamefont {Kallin},
  \citenamefont {Hastings}, \citenamefont {Melko},\ and\ \citenamefont
  {Singh}}]{Kallin11}%
  \BibitemOpen
  \bibfield  {author} {\bibinfo {author} {\bibfnamefont {A.~B.}\ \bibnamefont
  {Kallin}}, \bibinfo {author} {\bibfnamefont {M.~B.}\ \bibnamefont
  {Hastings}}, \bibinfo {author} {\bibfnamefont {R.~G.}\ \bibnamefont
  {Melko}},\ and\ \bibinfo {author} {\bibfnamefont {R.~R.~P.}\ \bibnamefont
  {Singh}},\ }\href {https://doi.org/10.1103/PhysRevB.84.165134} {\bibfield
  {journal} {\bibinfo  {journal} {Phys. Rev. B}\ }\textbf {\bibinfo {volume}
  {84}},\ \bibinfo {pages} {165134} (\bibinfo {year} {2011})}\BibitemShut
  {NoStop}%
\bibitem [{\citenamefont {Metlitski}\ and\ \citenamefont
  {Grover}(2011)}]{Metlitski11}%
  \BibitemOpen
  \bibfield  {author} {\bibinfo {author} {\bibfnamefont {M.~A.}\ \bibnamefont
  {Metlitski}}\ and\ \bibinfo {author} {\bibfnamefont {T.}~\bibnamefont
  {Grover}},\ }\href@noop {} {\bibinfo {title} {Entanglement entropy of systems
  with spontaneously broken continuous symmetry}} (\bibinfo {year} {2011}),\
  \Eprint {https://arxiv.org/abs/arXiv:1112.5166} {arXiv:1112.5166}
  \BibitemShut {NoStop}%
\bibitem [{\citenamefont {Laflorencie}\ \emph {et~al.}(2015)\citenamefont
  {Laflorencie}, \citenamefont {Luitz},\ and\ \citenamefont
  {Alet}}]{Laflorencie15}%
  \BibitemOpen
  \bibfield  {author} {\bibinfo {author} {\bibfnamefont {N.}~\bibnamefont
  {Laflorencie}}, \bibinfo {author} {\bibfnamefont {D.~J.}\ \bibnamefont
  {Luitz}},\ and\ \bibinfo {author} {\bibfnamefont {F.}~\bibnamefont {Alet}},\
  }\href {https://doi.org/10.1103/PhysRevB.92.115126} {\bibfield  {journal}
  {\bibinfo  {journal} {Phys. Rev. B}\ }\textbf {\bibinfo {volume} {92}},\
  \bibinfo {pages} {115126} (\bibinfo {year} {2015})}\BibitemShut {NoStop}%
\bibitem [{\citenamefont {Luitz}\ \emph {et~al.}(2015)\citenamefont {Luitz},
  \citenamefont {Plat}, \citenamefont {Alet},\ and\ \citenamefont
  {Laflorencie}}]{Luitz15ssb}%
  \BibitemOpen
  \bibfield  {author} {\bibinfo {author} {\bibfnamefont {D.~J.}\ \bibnamefont
  {Luitz}}, \bibinfo {author} {\bibfnamefont {X.}~\bibnamefont {Plat}},
  \bibinfo {author} {\bibfnamefont {F.}~\bibnamefont {Alet}},\ and\ \bibinfo
  {author} {\bibfnamefont {N.}~\bibnamefont {Laflorencie}},\ }\href
  {https://doi.org/10.1103/PhysRevB.91.155145} {\bibfield  {journal} {\bibinfo
  {journal} {Phys. Rev. B}\ }\textbf {\bibinfo {volume} {91}},\ \bibinfo
  {pages} {155145} (\bibinfo {year} {2015})}\BibitemShut {NoStop}%
\bibitem [{\citenamefont {{Alba}}\ \emph {et~al.}(2013)\citenamefont {{Alba}},
  \citenamefont {{Haque}},\ and\ \citenamefont {{L{\"a}uchli}}}]{Alba13}%
  \BibitemOpen
  \bibfield  {author} {\bibinfo {author} {\bibfnamefont {V.}~\bibnamefont
  {{Alba}}}, \bibinfo {author} {\bibfnamefont {M.}~\bibnamefont {{Haque}}},\
  and\ \bibinfo {author} {\bibfnamefont {A.~M.}\ \bibnamefont
  {{L{\"a}uchli}}},\ }\href {https://doi.org/10.1103/PhysRevLett.110.260403}
  {\bibfield  {journal} {\bibinfo  {journal} {Physical Review Letters}\
  }\textbf {\bibinfo {volume} {110}},\ \bibinfo {eid} {260403} (\bibinfo {year}
  {2013})}\BibitemShut {NoStop}%
\bibitem [{\citenamefont {Kolley}\ \emph {et~al.}(2013)\citenamefont {Kolley},
  \citenamefont {Depenbrock}, \citenamefont {McCulloch}, \citenamefont
  {Schollw\"ock},\ and\ \citenamefont {Alba}}]{Kolley13}%
  \BibitemOpen
  \bibfield  {author} {\bibinfo {author} {\bibfnamefont {F.}~\bibnamefont
  {Kolley}}, \bibinfo {author} {\bibfnamefont {S.}~\bibnamefont {Depenbrock}},
  \bibinfo {author} {\bibfnamefont {I.~P.}\ \bibnamefont {McCulloch}}, \bibinfo
  {author} {\bibfnamefont {U.}~\bibnamefont {Schollw\"ock}},\ and\ \bibinfo
  {author} {\bibfnamefont {V.}~\bibnamefont {Alba}},\ }\href
  {https://doi.org/10.1103/PhysRevB.88.144426} {\bibfield  {journal} {\bibinfo
  {journal} {Phys. Rev. B}\ }\textbf {\bibinfo {volume} {88}},\ \bibinfo
  {pages} {144426} (\bibinfo {year} {2013})}\BibitemShut {NoStop}%
\bibitem [{\citenamefont {Rademaker}(2015)}]{Rademaker15}%
  \BibitemOpen
  \bibfield  {author} {\bibinfo {author} {\bibfnamefont {L.}~\bibnamefont
  {Rademaker}},\ }\href {https://doi.org/10.1103/PhysRevB.92.144419} {\bibfield
   {journal} {\bibinfo  {journal} {Phys. Rev. B}\ }\textbf {\bibinfo {volume}
  {92}},\ \bibinfo {pages} {144419} (\bibinfo {year} {2015})}\BibitemShut
  {NoStop}%
\bibitem [{\citenamefont {Luitz}\ \emph {et~al.}(2014)\citenamefont {Luitz},
  \citenamefont {Alet},\ and\ \citenamefont {Laflorencie}}]{Luitz14}%
  \BibitemOpen
  \bibfield  {author} {\bibinfo {author} {\bibfnamefont {D.~J.}\ \bibnamefont
  {Luitz}}, \bibinfo {author} {\bibfnamefont {F.}~\bibnamefont {Alet}},\ and\
  \bibinfo {author} {\bibfnamefont {N.}~\bibnamefont {Laflorencie}},\ }\href
  {https://doi.org/10.1103/PhysRevLett.112.057203} {\bibfield  {journal}
  {\bibinfo  {journal} {Phys. Rev. Lett.}\ }\textbf {\bibinfo {volume} {112}},\
  \bibinfo {pages} {057203} (\bibinfo {year} {2014})}\BibitemShut {NoStop}%
\bibitem [{\citenamefont {Luitz}\ and\ \citenamefont
  {Laflorencie}(2017)}]{Luitz17}%
  \BibitemOpen
  \bibfield  {author} {\bibinfo {author} {\bibfnamefont {D.~J.}\ \bibnamefont
  {Luitz}}\ and\ \bibinfo {author} {\bibfnamefont {N.}~\bibnamefont
  {Laflorencie}},\ }\href {https://doi.org/10.21468/SciPostPhys.2.2.011}
  {\bibfield  {journal} {\bibinfo  {journal} {SciPost Phys.}\ }\textbf
  {\bibinfo {volume} {2}},\ \bibinfo {pages} {011} (\bibinfo {year}
  {2017})}\BibitemShut {NoStop}%
\bibitem [{\citenamefont {Misguich}\ \emph {et~al.}(2017)\citenamefont
  {Misguich}, \citenamefont {Pasquier},\ and\ \citenamefont
  {Oshikawa}}]{Misguich17}%
  \BibitemOpen
  \bibfield  {author} {\bibinfo {author} {\bibfnamefont {G.}~\bibnamefont
  {Misguich}}, \bibinfo {author} {\bibfnamefont {V.}~\bibnamefont {Pasquier}},\
  and\ \bibinfo {author} {\bibfnamefont {M.}~\bibnamefont {Oshikawa}},\ }\href
  {https://doi.org/10.1103/PhysRevB.95.195161} {\bibfield  {journal} {\bibinfo
  {journal} {Phys. Rev. B}\ }\textbf {\bibinfo {volume} {95}},\ \bibinfo
  {pages} {195161} (\bibinfo {year} {2017})}\BibitemShut {NoStop}%
\bibitem [{\citenamefont {Pethick}\ and\ \citenamefont
  {Smith}(2008)}]{Pethick_Smith_2008}%
  \BibitemOpen
  \bibfield  {author} {\bibinfo {author} {\bibfnamefont {C.~J.}\ \bibnamefont
  {Pethick}}\ and\ \bibinfo {author} {\bibfnamefont {H.}~\bibnamefont
  {Smith}},\ }\href {https://doi.org/10.1017/CBO9780511802850} {\emph {\bibinfo
  {title} {Bose--Einstein Condensation in Dilute Gases}}},\ \bibinfo {edition}
  {2nd}\ ed.\ (\bibinfo  {publisher} {Cambridge University Press},\ \bibinfo
  {year} {2008})\BibitemShut {NoStop}%
\bibitem [{\citenamefont {Malomed}(2008)}]{Malomed08}%
  \BibitemOpen
  \bibfield  {author} {\bibinfo {author} {\bibfnamefont {B.}~\bibnamefont
  {Malomed}},\ }\bibinfo {title} {Multi-component bose-einstein condensates:
  Theory},\ in\ \href {https://doi.org/10.1007/978-3-540-73591-5_15} {\emph
  {\bibinfo {booktitle} {Emergent Nonlinear Phenomena in Bose-Einstein
  Condensates: Theory and Experiment}}},\ \bibinfo {editor} {edited by\
  \bibinfo {editor} {\bibfnamefont {P.~G.}\ \bibnamefont {Kevrekidis}},
  \bibinfo {editor} {\bibfnamefont {D.~J.}\ \bibnamefont {Frantzeskakis}},\
  and\ \bibinfo {editor} {\bibfnamefont {R.}~\bibnamefont
  {Carretero-Gonz{\'a}lez}}}\ (\bibinfo  {publisher} {Springer Berlin
  Heidelberg},\ \bibinfo {address} {Berlin, Heidelberg},\ \bibinfo {year}
  {2008})\ pp.\ \bibinfo {pages} {287--305}\BibitemShut {NoStop}%
\bibitem [{\citenamefont {Hall}(2008)}]{Hall08}%
  \BibitemOpen
  \bibfield  {author} {\bibinfo {author} {\bibfnamefont {D.~S.}\ \bibnamefont
  {Hall}},\ }\bibinfo {title} {Multi-component condensates: Experiment},\ in\
  \href {https://doi.org/10.1007/978-3-540-73591-5_16} {\emph {\bibinfo
  {booktitle} {Emergent Nonlinear Phenomena in Bose-Einstein Condensates:
  Theory and Experiment}}},\ \bibinfo {editor} {edited by\ \bibinfo {editor}
  {\bibfnamefont {P.~G.}\ \bibnamefont {Kevrekidis}}, \bibinfo {editor}
  {\bibfnamefont {D.~J.}\ \bibnamefont {Frantzeskakis}},\ and\ \bibinfo
  {editor} {\bibfnamefont {R.}~\bibnamefont {Carretero-Gonz{\'a}lez}}}\
  (\bibinfo  {publisher} {Springer Berlin Heidelberg},\ \bibinfo {address}
  {Berlin, Heidelberg},\ \bibinfo {year} {2008})\ pp.\ \bibinfo {pages}
  {307--327}\BibitemShut {NoStop}%
\bibitem [{\citenamefont {Ho}\ and\ \citenamefont {Shenoy}(1996)}]{Ho96}%
  \BibitemOpen
  \bibfield  {author} {\bibinfo {author} {\bibfnamefont {T.-L.}\ \bibnamefont
  {Ho}}\ and\ \bibinfo {author} {\bibfnamefont {V.~B.}\ \bibnamefont
  {Shenoy}},\ }\href {https://doi.org/10.1103/PhysRevLett.77.3276} {\bibfield
  {journal} {\bibinfo  {journal} {Phys. Rev. Lett.}\ }\textbf {\bibinfo
  {volume} {77}},\ \bibinfo {pages} {3276} (\bibinfo {year}
  {1996})}\BibitemShut {NoStop}%
\bibitem [{\citenamefont {Ao}\ and\ \citenamefont {Chui}(1998)}]{Ao98}%
  \BibitemOpen
  \bibfield  {author} {\bibinfo {author} {\bibfnamefont {P.}~\bibnamefont
  {Ao}}\ and\ \bibinfo {author} {\bibfnamefont {S.~T.}\ \bibnamefont {Chui}},\
  }\href {https://doi.org/10.1103/PhysRevA.58.4836} {\bibfield  {journal}
  {\bibinfo  {journal} {Phys. Rev. A}\ }\textbf {\bibinfo {volume} {58}},\
  \bibinfo {pages} {4836} (\bibinfo {year} {1998})}\BibitemShut {NoStop}%
\bibitem [{\citenamefont {Timmermans}(1998)}]{Timmermans98}%
  \BibitemOpen
  \bibfield  {author} {\bibinfo {author} {\bibfnamefont {E.}~\bibnamefont
  {Timmermans}},\ }\href {https://doi.org/10.1103/PhysRevLett.81.5718}
  {\bibfield  {journal} {\bibinfo  {journal} {Phys. Rev. Lett.}\ }\textbf
  {\bibinfo {volume} {81}},\ \bibinfo {pages} {5718} (\bibinfo {year}
  {1998})}\BibitemShut {NoStop}%
\bibitem [{\citenamefont {Cazalilla}\ and\ \citenamefont
  {Ho}(2003)}]{Cazalilla03}%
  \BibitemOpen
  \bibfield  {author} {\bibinfo {author} {\bibfnamefont {M.~A.}\ \bibnamefont
  {Cazalilla}}\ and\ \bibinfo {author} {\bibfnamefont {A.~F.}\ \bibnamefont
  {Ho}},\ }\href {https://doi.org/10.1103/PhysRevLett.91.150403} {\bibfield
  {journal} {\bibinfo  {journal} {Phys. Rev. Lett.}\ }\textbf {\bibinfo
  {volume} {91}},\ \bibinfo {pages} {150403} (\bibinfo {year}
  {2003})}\BibitemShut {NoStop}%
\bibitem [{\citenamefont {Popov}(1972)}]{Popov72}%
  \BibitemOpen
  \bibfield  {author} {\bibinfo {author} {\bibfnamefont {V.~N.}\ \bibnamefont
  {Popov}},\ }\href {https://doi.org/10.1007/BF01028563} {\bibfield  {journal}
  {\bibinfo  {journal} {Theoretical and Mathematical Physics}\ }\textbf
  {\bibinfo {volume} {11}},\ \bibinfo {pages} {478} (\bibinfo {year}
  {1972})}\BibitemShut {NoStop}%
\bibitem [{\citenamefont {Liu}(1998)}]{Liu98}%
  \BibitemOpen
  \bibfield  {author} {\bibinfo {author} {\bibfnamefont {W.~V.}\ \bibnamefont
  {Liu}},\ }\href {https://doi.org/10.1142/S0217979298001241} {\bibfield
  {journal} {\bibinfo  {journal} {International Journal of Modern Physics B}\
  }\textbf {\bibinfo {volume} {12}},\ \bibinfo {pages} {2103} (\bibinfo {year}
  {1998})}\BibitemShut {NoStop}%
\bibitem [{\citenamefont {Al~Khawaja}\ \emph {et~al.}(2002)\citenamefont
  {Al~Khawaja}, \citenamefont {Andersen}, \citenamefont {Proukakis},\ and\
  \citenamefont {Stoof}}]{Khawaja02}%
  \BibitemOpen
  \bibfield  {author} {\bibinfo {author} {\bibfnamefont {U.}~\bibnamefont
  {Al~Khawaja}}, \bibinfo {author} {\bibfnamefont {J.~O.}\ \bibnamefont
  {Andersen}}, \bibinfo {author} {\bibfnamefont {N.~P.}\ \bibnamefont
  {Proukakis}},\ and\ \bibinfo {author} {\bibfnamefont {H.~T.~C.}\ \bibnamefont
  {Stoof}},\ }\href {https://doi.org/10.1103/PhysRevA.66.013615} {\bibfield
  {journal} {\bibinfo  {journal} {Phys. Rev. A}\ }\textbf {\bibinfo {volume}
  {66}},\ \bibinfo {pages} {013615} (\bibinfo {year} {2002})}\BibitemShut
  {NoStop}%
\bibitem [{\citenamefont {Mora}\ and\ \citenamefont {Castin}(2003)}]{Mora03}%
  \BibitemOpen
  \bibfield  {author} {\bibinfo {author} {\bibfnamefont {C.}~\bibnamefont
  {Mora}}\ and\ \bibinfo {author} {\bibfnamefont {Y.}~\bibnamefont {Castin}},\
  }\href {https://doi.org/10.1103/PhysRevA.67.053615} {\bibfield  {journal}
  {\bibinfo  {journal} {Phys. Rev. A}\ }\textbf {\bibinfo {volume} {67}},\
  \bibinfo {pages} {053615} (\bibinfo {year} {2003})}\BibitemShut {NoStop}%
\bibitem [{\citenamefont {Goldstein}\ and\ \citenamefont
  {Meystre}(1997)}]{Goldstein97}%
  \BibitemOpen
  \bibfield  {author} {\bibinfo {author} {\bibfnamefont {E.~V.}\ \bibnamefont
  {Goldstein}}\ and\ \bibinfo {author} {\bibfnamefont {P.}~\bibnamefont
  {Meystre}},\ }\href {https://doi.org/10.1103/PhysRevA.55.2935} {\bibfield
  {journal} {\bibinfo  {journal} {Phys. Rev. A}\ }\textbf {\bibinfo {volume}
  {55}},\ \bibinfo {pages} {2935} (\bibinfo {year} {1997})}\BibitemShut
  {NoStop}%
\bibitem [{\citenamefont {Search}\ \emph {et~al.}(2001)\citenamefont {Search},
  \citenamefont {Rojo},\ and\ \citenamefont {Berman}}]{Search01}%
  \BibitemOpen
  \bibfield  {author} {\bibinfo {author} {\bibfnamefont {C.~P.}\ \bibnamefont
  {Search}}, \bibinfo {author} {\bibfnamefont {A.~G.}\ \bibnamefont {Rojo}},\
  and\ \bibinfo {author} {\bibfnamefont {P.~R.}\ \bibnamefont {Berman}},\
  }\href {https://doi.org/10.1103/PhysRevA.64.013615} {\bibfield  {journal}
  {\bibinfo  {journal} {Phys. Rev. A}\ }\textbf {\bibinfo {volume} {64}},\
  \bibinfo {pages} {013615} (\bibinfo {year} {2001})}\BibitemShut {NoStop}%
\bibitem [{\citenamefont {Tommasini}\ \emph {et~al.}(2003)\citenamefont
  {Tommasini}, \citenamefont {de~Passos}, \citenamefont {de~Toledo~Piza},
  \citenamefont {Hussein},\ and\ \citenamefont {Timmermans}}]{Tommasini03}%
  \BibitemOpen
  \bibfield  {author} {\bibinfo {author} {\bibfnamefont {P.}~\bibnamefont
  {Tommasini}}, \bibinfo {author} {\bibfnamefont {E.~J.~V.}\ \bibnamefont
  {de~Passos}}, \bibinfo {author} {\bibfnamefont {A.~F.~R.}\ \bibnamefont
  {de~Toledo~Piza}}, \bibinfo {author} {\bibfnamefont {M.~S.}\ \bibnamefont
  {Hussein}},\ and\ \bibinfo {author} {\bibfnamefont {E.}~\bibnamefont
  {Timmermans}},\ }\href {https://doi.org/10.1103/PhysRevA.67.023606}
  {\bibfield  {journal} {\bibinfo  {journal} {Phys. Rev. A}\ }\textbf {\bibinfo
  {volume} {67}},\ \bibinfo {pages} {023606} (\bibinfo {year}
  {2003})}\BibitemShut {NoStop}%
\bibitem [{\citenamefont {Whitlock}\ and\ \citenamefont
  {Bouchoule}(2003)}]{Whitlock03}%
  \BibitemOpen
  \bibfield  {author} {\bibinfo {author} {\bibfnamefont {N.~K.}\ \bibnamefont
  {Whitlock}}\ and\ \bibinfo {author} {\bibfnamefont {I.}~\bibnamefont
  {Bouchoule}},\ }\href {https://doi.org/10.1103/PhysRevA.68.053609} {\bibfield
   {journal} {\bibinfo  {journal} {Phys. Rev. A}\ }\textbf {\bibinfo {volume}
  {68}},\ \bibinfo {pages} {053609} (\bibinfo {year} {2003})}\BibitemShut
  {NoStop}%
\bibitem [{\citenamefont {Tononi}\ \emph {et~al.}(2020)\citenamefont {Tononi},
  \citenamefont {Toigo}, \citenamefont {Wimberger}, \citenamefont
  {Cappellaro},\ and\ \citenamefont {Salasnich}}]{Tononi20}%
  \BibitemOpen
  \bibfield  {author} {\bibinfo {author} {\bibfnamefont {A.}~\bibnamefont
  {Tononi}}, \bibinfo {author} {\bibfnamefont {F.}~\bibnamefont {Toigo}},
  \bibinfo {author} {\bibfnamefont {S.}~\bibnamefont {Wimberger}}, \bibinfo
  {author} {\bibfnamefont {A.}~\bibnamefont {Cappellaro}},\ and\ \bibinfo
  {author} {\bibfnamefont {L.}~\bibnamefont {Salasnich}},\ }\href
  {https://doi.org/10.1088/1367-2630/ab965d} {\bibfield  {journal} {\bibinfo
  {journal} {New Journal of Physics}\ }\textbf {\bibinfo {volume} {22}},\
  \bibinfo {pages} {073020} (\bibinfo {year} {2020})}\BibitemShut {NoStop}%
\bibitem [{\citenamefont {Leggett}\ and\ \citenamefont
  {Sols}(1991)}]{Leggett91}%
  \BibitemOpen
  \bibfield  {author} {\bibinfo {author} {\bibfnamefont {A.~J.}\ \bibnamefont
  {Leggett}}\ and\ \bibinfo {author} {\bibfnamefont {F.}~\bibnamefont {Sols}},\
  }\href {https://doi.org/10.1007/BF01883640} {\bibfield  {journal} {\bibinfo
  {journal} {Foundations of Physics}\ }\textbf {\bibinfo {volume} {21}},\
  \bibinfo {pages} {353} (\bibinfo {year} {1991})}\BibitemShut {NoStop}%
\bibitem [{\citenamefont {Tasaki}(2020)}]{Tasaki20}%
  \BibitemOpen
  \bibfield  {author} {\bibinfo {author} {\bibfnamefont {H.}~\bibnamefont
  {Tasaki}},\ }\href {https://doi.org/10.1007/s10955-019-02435-9} {\bibfield
  {journal} {\bibinfo  {journal} {Journal of Statistical Physics}\ }\textbf
  {\bibinfo {volume} {178}},\ \bibinfo {pages} {379} (\bibinfo {year}
  {2020})}\BibitemShut {NoStop}%
\bibitem [{\citenamefont {Foini}\ and\ \citenamefont
  {Giamarchi}(2015)}]{Foini15}%
  \BibitemOpen
  \bibfield  {author} {\bibinfo {author} {\bibfnamefont {L.}~\bibnamefont
  {Foini}}\ and\ \bibinfo {author} {\bibfnamefont {T.}~\bibnamefont
  {Giamarchi}},\ }\href {https://doi.org/10.1103/PhysRevA.91.023627} {\bibfield
   {journal} {\bibinfo  {journal} {Phys. Rev. A}\ }\textbf {\bibinfo {volume}
  {91}},\ \bibinfo {pages} {023627} (\bibinfo {year} {2015})}\BibitemShut
  {NoStop}%
\bibitem [{\citenamefont {Foini}(2017)}]{Foini17}%
  \BibitemOpen
  \bibfield  {author} {\bibinfo {author} {\bibfnamefont {T.}~\bibnamefont
  {Foini}, \bibfnamefont {L.and~Giamarchi}},\ }\href
  {https://doi.org/10.1140/epjst/e2016-60383-x} {\bibfield  {journal} {\bibinfo
   {journal} {The European Physical Journal Special Topics}\ }\textbf {\bibinfo
  {volume} {226}},\ \bibinfo {pages} {2763} (\bibinfo {year}
  {2017})}\BibitemShut {NoStop}%
\bibitem [{Note1()}]{Note1}%
  \BibitemOpen
  \bibinfo {note} {Owing to the compactness $\theta _\alpha \equiv \theta
  _\alpha + 2\pi $, a winding term $2\pi \protect \bm {M}\cdot {\protect
  \mathbf {r}}/L$ with $\protect \bm {M}\in \protect \mathbb {Z}^d$ is also
  allowed in the phase $\theta _\alpha ({\protect \mathbf {r}})$, as is often
  discussed in the TLL description \cite {Cazalilla04}. Here, we do not include
  such a term as it vanishes in the ground state.}\BibitemShut {Stop}%
\bibitem [{\citenamefont {Peschel}(2003)}]{Peschel03}%
  \BibitemOpen
  \bibfield  {author} {\bibinfo {author} {\bibfnamefont {I.}~\bibnamefont
  {Peschel}},\ }\href {https://doi.org/10.1088/0305-4470/36/14/101} {\bibfield
  {journal} {\bibinfo  {journal} {J. Physics A: Mathematical and General}\
  }\textbf {\bibinfo {volume} {36}},\ \bibinfo {pages} {L205} (\bibinfo {year}
  {2003})}\BibitemShut {NoStop}%
\bibitem [{\citenamefont {Peschel}\ and\ \citenamefont
  {Eisler}(2009)}]{Peschel09}%
  \BibitemOpen
  \bibfield  {author} {\bibinfo {author} {\bibfnamefont {I.}~\bibnamefont
  {Peschel}}\ and\ \bibinfo {author} {\bibfnamefont {V.}~\bibnamefont
  {Eisler}},\ }\href {https://doi.org/10.1088/1751-8113/42/50/504003}
  {\bibfield  {journal} {\bibinfo  {journal} {Journal of Physics A:
  Mathematical and Theoretical}\ }\textbf {\bibinfo {volume} {42}},\ \bibinfo
  {pages} {504003} (\bibinfo {year} {2009})}\BibitemShut {NoStop}%
\bibitem [{\citenamefont {Cazalilla}(2004)}]{Cazalilla04}%
  \BibitemOpen
  \bibfield  {author} {\bibinfo {author} {\bibfnamefont {M.~A.}\ \bibnamefont
  {Cazalilla}},\ }\href {https://doi.org/10.1088/0953-4075/37/7/051} {\bibfield
   {journal} {\bibinfo  {journal} {Journal of Physics B: Atomic, Molecular and
  Optical Physics}\ }\textbf {\bibinfo {volume} {37}},\ \bibinfo {pages} {S1}
  (\bibinfo {year} {2004})}\BibitemShut {NoStop}%
\bibitem [{\citenamefont {Fradkin}\ \emph {et~al.}(1993)\citenamefont
  {Fradkin}, \citenamefont {Moreno},\ and\ \citenamefont
  {Schaposnik}}]{Fradkin93}%
  \BibitemOpen
  \bibfield  {author} {\bibinfo {author} {\bibfnamefont {E.}~\bibnamefont
  {Fradkin}}, \bibinfo {author} {\bibfnamefont {E.}~\bibnamefont {Moreno}},\
  and\ \bibinfo {author} {\bibfnamefont {F.~A.}\ \bibnamefont {Schaposnik}},\
  }\href {https://doi.org/http://dx.doi.org/10.1016/0550-3213(93)90521-P}
  {\bibfield  {journal} {\bibinfo  {journal} {Nuclear Physics B}\ }\textbf
  {\bibinfo {volume} {392}},\ \bibinfo {pages} {667 } (\bibinfo {year}
  {1993})}\BibitemShut {NoStop}%
\bibitem [{\citenamefont {St\'ephan}\ \emph {et~al.}(2009)\citenamefont
  {St\'ephan}, \citenamefont {Furukawa}, \citenamefont {Misguich},\ and\
  \citenamefont {Pasquier}}]{Stephan09}%
  \BibitemOpen
  \bibfield  {author} {\bibinfo {author} {\bibfnamefont {J.-M.}\ \bibnamefont
  {St\'ephan}}, \bibinfo {author} {\bibfnamefont {S.}~\bibnamefont {Furukawa}},
  \bibinfo {author} {\bibfnamefont {G.}~\bibnamefont {Misguich}},\ and\
  \bibinfo {author} {\bibfnamefont {V.}~\bibnamefont {Pasquier}},\ }\href
  {https://doi.org/10.1103/PhysRevB.80.184421} {\bibfield  {journal} {\bibinfo
  {journal} {Phys. Rev. B}\ }\textbf {\bibinfo {volume} {80}},\ \bibinfo
  {pages} {184421} (\bibinfo {year} {2009})}\BibitemShut {NoStop}%
\bibitem [{\citenamefont {Penrose}\ and\ \citenamefont
  {Onsager}(1956)}]{Penrose56}%
  \BibitemOpen
  \bibfield  {author} {\bibinfo {author} {\bibfnamefont {O.}~\bibnamefont
  {Penrose}}\ and\ \bibinfo {author} {\bibfnamefont {L.}~\bibnamefont
  {Onsager}},\ }\href {https://doi.org/10.1103/PhysRev.104.576} {\bibfield
  {journal} {\bibinfo  {journal} {Phys. Rev.}\ }\textbf {\bibinfo {volume}
  {104}},\ \bibinfo {pages} {576} (\bibinfo {year} {1956})}\BibitemShut
  {NoStop}%
\bibitem [{\citenamefont {Yang}(1962)}]{Yang62}%
  \BibitemOpen
  \bibfield  {author} {\bibinfo {author} {\bibfnamefont {C.~N.}\ \bibnamefont
  {Yang}},\ }\href {https://doi.org/10.1103/RevModPhys.34.694} {\bibfield
  {journal} {\bibinfo  {journal} {Rev. Mod. Phys.}\ }\textbf {\bibinfo {volume}
  {34}},\ \bibinfo {pages} {694} (\bibinfo {year} {1962})}\BibitemShut
  {NoStop}%
\bibitem [{Note2()}]{Note2}%
  \BibitemOpen
  \bibinfo {note} {The power-law decay in terms of the chord distance as in
  Eq.\ \protect \textup {\hbox {\mathsurround \z@ \protect \normalfont
  (\ignorespaces \ref {eq:corr_psi_d1}\unskip \@@italiccorr )}} is well-known
  in the behavior of correlation functions of primary fields in conformal field
  theory. See, e.g., Appendix C of Ref.\ \cite {Cazalilla04}.}\BibitemShut
  {Stop}%
\bibitem [{\citenamefont {Song}\ \emph {et~al.}(2010)\citenamefont {Song},
  \citenamefont {Rachel},\ and\ \citenamefont {Le~Hur}}]{Song10}%
  \BibitemOpen
  \bibfield  {author} {\bibinfo {author} {\bibfnamefont {H.~F.}\ \bibnamefont
  {Song}}, \bibinfo {author} {\bibfnamefont {S.}~\bibnamefont {Rachel}},\ and\
  \bibinfo {author} {\bibfnamefont {K.}~\bibnamefont {Le~Hur}},\ }\href
  {https://doi.org/10.1103/PhysRevB.82.012405} {\bibfield  {journal} {\bibinfo
  {journal} {Phys. Rev. B}\ }\textbf {\bibinfo {volume} {82}},\ \bibinfo
  {pages} {012405} (\bibinfo {year} {2010})}\BibitemShut {NoStop}%
\bibitem [{\citenamefont {Berg}\ \emph {et~al.}(2008)\citenamefont {Berg},
  \citenamefont {Dalla~Torre}, \citenamefont {Giamarchi},\ and\ \citenamefont
  {Altman}}]{Berg08}%
  \BibitemOpen
  \bibfield  {author} {\bibinfo {author} {\bibfnamefont {E.}~\bibnamefont
  {Berg}}, \bibinfo {author} {\bibfnamefont {E.~G.}\ \bibnamefont
  {Dalla~Torre}}, \bibinfo {author} {\bibfnamefont {T.}~\bibnamefont
  {Giamarchi}},\ and\ \bibinfo {author} {\bibfnamefont {E.}~\bibnamefont
  {Altman}},\ }\href {https://doi.org/10.1103/PhysRevB.77.245119} {\bibfield
  {journal} {\bibinfo  {journal} {Phys. Rev. B}\ }\textbf {\bibinfo {volume}
  {77}},\ \bibinfo {pages} {245119} (\bibinfo {year} {2008})}\BibitemShut
  {NoStop}%
\bibitem [{\citenamefont {Endres}\ \emph {et~al.}(2011)\citenamefont {Endres},
  \citenamefont {Cheneau}, \citenamefont {Fukuhara}, \citenamefont
  {Weitenberg}, \citenamefont {Schau{\ss}}, \citenamefont {Gross},
  \citenamefont {Mazza}, \citenamefont {Ba{\~n}uls}, \citenamefont {Pollet},
  \citenamefont {Bloch},\ and\ \citenamefont {Kuhr}}]{Endres11}%
  \BibitemOpen
  \bibfield  {author} {\bibinfo {author} {\bibfnamefont {M.}~\bibnamefont
  {Endres}}, \bibinfo {author} {\bibfnamefont {M.}~\bibnamefont {Cheneau}},
  \bibinfo {author} {\bibfnamefont {T.}~\bibnamefont {Fukuhara}}, \bibinfo
  {author} {\bibfnamefont {C.}~\bibnamefont {Weitenberg}}, \bibinfo {author}
  {\bibfnamefont {P.}~\bibnamefont {Schau{\ss}}}, \bibinfo {author}
  {\bibfnamefont {C.}~\bibnamefont {Gross}}, \bibinfo {author} {\bibfnamefont
  {L.}~\bibnamefont {Mazza}}, \bibinfo {author} {\bibfnamefont {M.~C.}\
  \bibnamefont {Ba{\~n}uls}}, \bibinfo {author} {\bibfnamefont
  {L.}~\bibnamefont {Pollet}}, \bibinfo {author} {\bibfnamefont
  {I.}~\bibnamefont {Bloch}},\ and\ \bibinfo {author} {\bibfnamefont
  {S.}~\bibnamefont {Kuhr}},\ }\href {https://doi.org/10.1126/science.1209284}
  {\bibfield  {journal} {\bibinfo  {journal} {Science}\ }\textbf {\bibinfo
  {volume} {334}},\ \bibinfo {pages} {200} (\bibinfo {year}
  {2011})}\BibitemShut {NoStop}%
\bibitem [{\citenamefont {Kawaguchi}\ and\ \citenamefont
  {Ueda}(2012)}]{Kawaguchi12}%
  \BibitemOpen
  \bibfield  {author} {\bibinfo {author} {\bibfnamefont {Y.}~\bibnamefont
  {Kawaguchi}}\ and\ \bibinfo {author} {\bibfnamefont {M.}~\bibnamefont
  {Ueda}},\ }\href
  {https://doi.org/https://doi.org/10.1016/j.physrep.2012.07.005} {\bibfield
  {journal} {\bibinfo  {journal} {Physics Reports}\ }\textbf {\bibinfo {volume}
  {520}},\ \bibinfo {pages} {253 } (\bibinfo {year} {2012})},\ \bibinfo {note}
  {spinor Bose--Einstein condensates}\BibitemShut {NoStop}%
\bibitem [{\citenamefont {Lin}\ \emph {et~al.}(2011)\citenamefont {Lin},
  \citenamefont {Jim{\'e}nez-Garc{\'\i}a},\ and\ \citenamefont
  {Spielman}}]{Lin11}%
  \BibitemOpen
  \bibfield  {author} {\bibinfo {author} {\bibfnamefont {Y.~J.}\ \bibnamefont
  {Lin}}, \bibinfo {author} {\bibfnamefont {K.}~\bibnamefont
  {Jim{\'e}nez-Garc{\'\i}a}},\ and\ \bibinfo {author} {\bibfnamefont {I.~B.}\
  \bibnamefont {Spielman}},\ }\href {http://dx.doi.org/10.1038/nature09887}
  {\bibfield  {journal} {\bibinfo  {journal} {Nature}\ }\textbf {\bibinfo
  {volume} {471}},\ \bibinfo {pages} {83} (\bibinfo {year} {2011})}\BibitemShut
  {NoStop}%
\bibitem [{\citenamefont {Zhai}(2012)}]{Zhai12_review}%
  \BibitemOpen
  \bibfield  {author} {\bibinfo {author} {\bibfnamefont {H.}~\bibnamefont
  {Zhai}},\ }\href {https://doi.org/10.1142/S0217979212300010} {\bibfield
  {journal} {\bibinfo  {journal} {International Journal of Modern Physics B}\
  }\textbf {\bibinfo {volume} {26}},\ \bibinfo {pages} {1230001} (\bibinfo
  {year} {2012})}\BibitemShut {NoStop}%
\bibitem [{\citenamefont {Po}\ \emph {et~al.}(2014)\citenamefont {Po},
  \citenamefont {Chen},\ and\ \citenamefont {Zhou}}]{Po14}%
  \BibitemOpen
  \bibfield  {author} {\bibinfo {author} {\bibfnamefont {H.~C.}\ \bibnamefont
  {Po}}, \bibinfo {author} {\bibfnamefont {W.}~\bibnamefont {Chen}},\ and\
  \bibinfo {author} {\bibfnamefont {Q.}~\bibnamefont {Zhou}},\ }\href
  {https://doi.org/10.1103/PhysRevA.90.011602} {\bibfield  {journal} {\bibinfo
  {journal} {Phys. Rev. A}\ }\textbf {\bibinfo {volume} {90}},\ \bibinfo
  {pages} {011602(R)} (\bibinfo {year} {2014})}\BibitemShut {NoStop}%
\bibitem [{\citenamefont {Yoshino}\ \emph {et~al.}(2019)\citenamefont
  {Yoshino}, \citenamefont {Furukawa}, \citenamefont {Higashikawa},\ and\
  \citenamefont {Ueda}}]{Yoshino19}%
  \BibitemOpen
  \bibfield  {author} {\bibinfo {author} {\bibfnamefont {T.}~\bibnamefont
  {Yoshino}}, \bibinfo {author} {\bibfnamefont {S.}~\bibnamefont {Furukawa}},
  \bibinfo {author} {\bibfnamefont {S.}~\bibnamefont {Higashikawa}},\ and\
  \bibinfo {author} {\bibfnamefont {M.}~\bibnamefont {Ueda}},\ }\href
  {https://doi.org/10.1088/1367-2630/aaf373} {\bibfield  {journal} {\bibinfo
  {journal} {New Journal of Physics}\ }\textbf {\bibinfo {volume} {21}},\
  \bibinfo {pages} {015001} (\bibinfo {year} {2019})}\BibitemShut {NoStop}%
\bibitem [{\citenamefont {Beekman}(2020)}]{Beekman20}%
  \BibitemOpen
  \bibfield  {author} {\bibinfo {author} {\bibfnamefont {A.~J.}\ \bibnamefont
  {Beekman}},\ }\bibfield  {journal} {\bibinfo  {journal} {Progress of
  Theoretical and Experimental Physics}\ }\textbf {\bibinfo {volume} {2020}},\
  \href {https://doi.org/10.1093/ptep/ptaa088} {10.1093/ptep/ptaa088} (\bibinfo
  {year} {2020}),\ \bibinfo {note} {073B09},\ \Eprint
  {https://arxiv.org/abs/https://academic.oup.com/ptep/article-pdf/2020/7/073B09/33535191/ptaa088.pdf}
  {https://academic.oup.com/ptep/article-pdf/2020/7/073B09/33535191/ptaa088.pdf}
  \BibitemShut {NoStop}%
\end{thebibliography}%

\end{document}